# Non-invasive assessment by B-mode ultrasound of arterial pulse wave intensity and its modification during ventricular dysfunction

Ethan M. Rowland, Kai Riemer, Kevin Lichtenstein, Mengxing Tang, Peter D. Weinberg *

*Department of Bioengineering, Imperial College London, London, UK*

*Corresponding Author
Professor Peter D. Weinberg
Department of Bioengineering,
Imperial College London,
London SW7 2AZ, UK
p.weinberg@imperial.ac.uk

Running title: Non-invasive measurement of wave intensity

## Abstract

Arterial pulse waves contain clinically useful information: their intensity varies with cardiac performance, their speed (“pulse wave velocity;” PWV) depends on arterial stiffness and their reflection is affected by conduit artery tone. Here we demonstrate a novel method for non-invasively assessing wave properties. The analysis was based on changes in blood flow velocity and arterial wall diameter during the cardiac cycle. Velocity and diameter were determined by tracking the movement of speckles in successive B-mode ultrasound images acquired at high temporal frequency with an ultrafast plane-wave scanner. Blood speckle was detected in the absence of contrast agents by using singular value decomposition, and it was processed by cross-correlation techniques that correct biases in ultrasound imaging velocimetry. Results obtained in the rabbit aorta were compared with a conventional analysis based on blood velocity and pressure, employing measurements obtained with an intra-arterial catheter. The catheter-based measurements had a poorer frequency response and greater lags but patterns of forward and backward travelling waves were consistent between the conventional and new methods. Errors in PWV were also similar in magnitude, although opposite in direction. Comparable reductions in wave intensity and delays in wave arrival were detected by the two methods when ventricular dysfunction was induced pharmacologically. The non-invasive method was applied to the carotid artery of a healthy human subject and gave a PWV and patterns of wave intensity that were consistent with earlier measurements. The new system may have clinical utility, for example in screening for, and monitoring of, heart failure.

**Keywords:** wave intensity analysis; speckle tracking; UIV; esmolol

## Introduction

At the start of systole, ventricular contraction causes a wave of increased blood pressure, blood velocity and vessel diameter that propagates along the systemic arteries [1]. At the end of systole, when contraction slows and then reverses, a wave of decreased pressure, velocity and diameter propagates in the same direction [2]. These waves partially reflect and re-reflect where vessel geometry or structure changes [3]. They carry information concerning cardiac and vascular properties: wave speed (often termed pulse wave velocity; PWV) is related to arterial stiffness [4], wave reflections can be altered when vessel tone changes [5, 6], and wave intensity is determined by cardiac performance [7, 8]. Here we focus on the latter.

Curtis et al [7] found that the energy of the first wave was markedly reduced in stable compensated systolic heart failure and that the reduction increased with increasing severity on the NYHA scale. Sugawara et al [8] subsequently found that the first wave was reduced by approximately 50% in patients with dilated cardiomyopathy (i.e. impaired cardiac contractility), without significant change in the second wave; conversely, the second wave was reduced by approximately 50% in patients with hypertrophic cardiomyopathy (i.e. impaired diastolic function), without significant change in the first.

Both studies employed the methods of analysis developed by Parker and colleagues in which wave intensity, dI, is calculated as the product of dP and dU – respectively, the change in blood pressure (P) and velocity (U) over a short time interval in the artery of interest [9]. If the wave speed is also known (and it can be derived from dP and dU), the waves can additionally be divided into their forward- and backward-travelling components [10]. Measurements of P and U should be temporally and spatially coincident, and they need to be made with high temporal resolution to capture the rapid changes in wave intensity that occur over the cardiac cycle.

These requirements, and particularly the need for high-frequency pressure recordings, are problematic. Curtis et al [7] used Doppler ultrasound to obtain U and tonometry to obtain P, both in the carotid artery. Tonometry requires complex calibration to give true pressures, and the measurements of U and P cannot be made simultaneously. Suguwara et al [8] used Doppler ultrasound to measure U and M-mode ultrasound to measure arterial diameter, D, and assumed that D is proportional to P. This ignores the well-established nonlinearity of arterial stress-strain curves (see below). A technique which avoids these problems is to measure P and U with an intra-arterial catheter that has both a pressure transducer and a Doppler probe at its tip [11]. However, the invasiveness of the method limits its clinical utility.

Feng & Khir [12] developed a formulation in which wave intensity can be derived directly from U and D. A subsequent formulation by Biglino et al [13] employs flow rate, Q, and cross-sectional area, A. The intensities are not identical to those obtained from U and P – even the dimensions are different – but these systems are internally self-consistent. We [14] and others [15] have shown by numerical modelling that the formulation of Feng and Khir should have equivalent clinical utility to the more established method.

The advantage of these new systems is that U and D, or Q and A, can be obtained by non-invasive techniques. Pomella et al [16, 17] and Kowalski et al [18] used Doppler ultrasound to obtain U and B- or M-mode ultrasound to obtain D; Li et al [19] used MRI to obtain U and D; and Biglino and colleagues [e.g. 13, 20] used MRI to obtain Q and A. Useful wave intensity data were obtained in each case, but limitations are the well-established inaccuracies in Doppler velocity measurements [21], the difference in optimal beam angles for Doppler and B- or M-mode imaging, and the high cost and long acquisition time of MRI.

In the present work, we describe a non-invasive method in which D and U are both obtained from the same B-mode ultrasound images; it uses singular value decomposition (SVD) to separate the weak blood and strong tissue signals, cross-correlation between images to track the moving wall and blood speckles, and an ultrafast plane-wave scanner to adequately resolve the rapid acceleration and deceleration of blood during systole [22]. The method is validated by comparison with data obtained using the invasive catheter-based method and is shown to be capable of detecting ventricular dysfunction in rabbits. Preliminary data demonstrate that the technique can be translated to human subjects.

## Methods

### *Animal preparation*

All experiments complied with the Animals (Scientific Procedures) Act 1986 and were approved by the Animal Welfare and Ethical Review Body of Imperial College London. Twelve New Zealand White rabbits (HSDIF strain, Envigo, UK; age 81±13 days, weight 2.39±0.38 kg, mean±SD) were maintained on a 12:12 h light:dark cycle at 18 °C and fed a normal laboratory diet and tap water *ad libitum*. Each animal was pre-medicated with acepromazine (0.5 mg/kg *im*), anaesthetised with fentanyl fluanisone (Hypnorm, 0.3 mL/kg *im*) and midazolam (Hypnovel, 0.1 mL/kg *iv*), tracheotomised and ventilated (40 breaths/min, 45 cm $H_2O$ peak inflation pressure; Harvard Small Animal Ventilator). Anaesthesia was maintained with fentanyl fluanisone (0.1 mL/kg) and midazolam (0.1 mL/kg) approx. every 45 mins. Body temperature was monitored using a rectal probe and maintained with a heated pad. Blood oxygen saturation was monitored by pulse oximetry.

### *Ultrasound imaging*

A Vantage 64 LE ultrasound research platform (Verasonics, USA) equipped with an L11-4 probe was used to collect high frame rate images of the abdominal aorta distal to the renal arteries. The rabbit was tipped from its supine position to elevate its right side, and the probe angled in the opposite direction and clamped to give a stable, approximately coronal view through the aortic centreline. 64 elements located at the centre of the probe were used to transmit and receive giving a lateral field of view of 19.2 mm.

B-mode imaging was performed using a coherent plane wave compounding scheme [PMID: 19411209; PMID: 32130594]. A pulse sequence consisting of 3 plane waves (8Mhz, 1-cycle) steered from -5º to 5º in 5º steps was transmitted at a pulse repetition frequency of 3 kHz for an imaging depth of 20 mm. Successive pulse sequences were separated to give an effective frame rate of 1kHz after coherent summation of the 3 angled plane waves. Images were acquired for 2s. The backscattered radio-frequency (RF) signals were sampled at 25 MHz; echoes of each transmission were recorded for post-processing offline.

A 3-lead ECG signal was recorded concurrently via a PowerLab 26T data acquisition system (AD instruments, UK) connected to the Verasonics host computer running LabChart software. An analog trigger signal sent at the start of the image acquisition sequence from the Verasonics to the PowerLab enabled later time alignment of the arterial waveforms and ECG trace.

### *Intra-arterial measurements*

The proximal right femoral artery was exposed, lidocaine was administered topically (typically 3 sites, < 1mL total) and the femoral nerve was severed.

A Volcano ComboWire XT wire with pressure and Doppler sensors at the tip was connected to a ComboMap system (Phillips, USA), inserted into the artery and advanced into the abdominal aorta, just distal to the imaging site. The wire was rotated until the strongest Doppler signal was obtained. Within the ComboMap system, the wall filter was set at 400 Hz and the signal-to-noise threshold was adjusted manually to optimise the real-time Doppler envelope tracking.

The analog pressure and velocity output ports were connected to two analog inputs on the PowerLab data acquisition system (1v = 100 mmHg for pressure, 0.5 m/s for velocity, sample rate 1kHz). The Doppler sampling rate on the ComboMap system is 120 Hz.

### *Experimental protocol*

Imaging was performed at MI = 0.1 MPa. Once the catheter was inserted, at least three datasets were acquired with the image-based (Verasonics) and catheter-based (Volcano) systems. The catheter-mounted flow sensor was disconnected during imaging as it caused interference and the two techniques were therefore used alternately. Catheter-derived pressure data could, however, be recorded during the ultrasound imaging.

To assess effects of ventricular dysfunction, esmolol was administered to 10 rabbits; each received three boluses of increasing dose (1.5, 3.0 and 6.0 mg/kg, *iv*) at 10-minute intervals. Successive datasets were acquired with both systems every minute, starting just before the first dose. Two rabbits were given an equivalent volume of saline *iv* as a control.

### *Post-processing of invasive pressure data*

All post-processing was performed in Matlab R2020b (The MathWorks, USA).

Signal processing in the Volcano Combowire pressure-measuring system introduces a delay. To compensate for this, the pressure traces were brought forward in time by 20 ms, a value determined by aligning the peak pressure determined using the Volcano system with the peak determined using a high-fidelity catheter system simultaneously in an elastic tube model *in vitro* (Supplementary Figure 1).

This shift was increased by an additional 0.5-5 ms to account for the rabbit-specific displacement of the Combowire tip from the centre of the B-mode field of view used to determine D and U non-invasively; the distance was converted to a time delay using the rabbit-specific wave speed.

Even after peak alignment, the foot of the Volcano-derived pressure waveform was less acute and occurred ~5 ms earlier than the foot in the high-fidelity catheter-derived waveform when both were employed in the *in vitro* system (Supplementary Figure 1). This difference was also observed when comparing Volcano-derived pressure with diameter measured by the non-invasive ultrasound method. Analysis of the power spectra revealed low-order, frequency-based filtering in the Volcano system. A filter with comparable characteristics was constructed; it was applied to measurements of D and U when processing the data with PU and PD methods. This adjustment makes data from the different sources directly comparable and avoids anomalous behaviour such as non-linear slopes of the P-D and P-U loops in early systole, but does not have a substantial effect on mean wave speeds or intensities (Supplementary Figure 2).

### *Post-processing of non-invasive ultrasound data*

Velocimetry methods are based on earlier studies and validations by Riemer et al [22, 23]. Post-processing was again performed in Matlab R2020b.

*Beamforming and compounding*

The RF channel data were reconstructed using delay-and-sum beamforming (assuming a 1540 m/s speed of sound for delay calculations) and then Hilbert transformed. Echoes from the three angled plane waves were coherently summed resulting in a final ensemble image size of $N_x$ = 128, $N_z$ = 416 and $N_t$ = 2000 (lateral pixel size 0.15 mm; axial pixel size 0.048 mm).

*SVD filtering*

SVD filtering was used to separate the blood signal from the tissue signal and noise (see [24]). The image was cropped so that its axial dimension was slightly larger than the vessel diameter. Then the ensemble image S was reshaped into a 2D spatiotemporal representation (size Nx.Nz x Nt) and decomposed as follows:

$$S(xz, t) = \sum_{i=1}^{N_t} \lambda_i \, u_i(xz) \, v_i(t)^T$$

where $\lambda_i$ are the ordered singular values of $S$, and $u_i$ and $v_i$ are the corresponding spatial and temporal singular vectors. Assuming that tissue, blood and noise have different spatiotemporal characteristics – the first (and largest) singular values typically correspond to tissue, the next to blood, and the smallest to noise. The decomposition can be written as:

$$S(xz, t) = S_{tissue} + S_{blood} + S_{noise}$$

$$S(xz, t) = \sum_{i=1}^{Th_1} \lambda_i \, u_i(xz) \, v_i(t)^T + \sum_{i=Th_1+1}^{Th_2} \lambda_i \, u_i(xz) \, v_i(t)^T + \sum_{i=Th_2+1}^{N_t} \lambda_i \, u_i(xz) \, v_i(t)^T$$

where $Th_1$ and $Th_2$ are singular value thresholds. These were selected manually based on visual appearance of $u_i$ and the frequency of the temporal vectors $v_i$ (25). Finally, images were resampled using Matlab's griddedInterpolant() function to obtain isotropic pixel dimensions.

*Diameter measurement*

Diameter waveforms were computed by 1D cross-correlation of successive frames in $S_{tissue}$. First a ROI was selected on the anterior wall of the aorta, with a width equal to 5 A-lines and a height of 30 pixels. The signal values within the ROI were averaged laterally, and the location of the maximum corresponding to the inner layers of the wall was identified from the envelope. This was further refined by subpixel gaussian fitting.

1D cross-correlation functions were computed for each lateral location, according to the equation below, with $a^k$ and $b^{k+1}$ vectors of equal size. (The wall moved little so a larger search window was unnecessary.) The resulting correlation functions were averaged and the maximum was identified to give the axial displacement. Subpixel resolution was achieved by Gaussian fitting. This was repeated for the remaining image pairs, with the window offset by the previous displacement each time. The wall motion waveform was given by cumulatively summing the displacements over time.

The above was repeated for the posterior artery wall. The diameter change waveform was then given by the difference between the anterior and posterior waveforms, and addition of the initial diameter measured in frame 1 gave absolute diameter.

*Velocity measurement*

Velocity was measured by tracking the movement of blood-speckle patterns between frames in $S_{blood}$, using ultrasound imaging velocimetry. This can produce 2D velocity fields that are independent of beam angle. Briefly, sequential image pairs were divided into "interrogation windows" and the best cross-correlation for an interrogation window in the first image of the pair out of all those obtained with interrogation windows in the second image of the pair was used to show how far the scatterers in the interrogation window had moved in aggregate. Local velocity vectors were then obtained using the imaging frame rate and pixel resolution.

Correlation was Fourier-based for computational efficiency. Consider a template window $a^{(k)}$ and search window $b^{(k+1)}$ where $k$ is the frame number. Cross-correlation in the spatial-domain is equivalent to convolution in the frequency-domain:

$$r(\delta x, \delta z) = \mathcal{F}^{-1}\{A^{(k)}(\delta x, \delta z) \cdot B^{(k+1)*}(\delta x, \delta z)\}$$

where $A$ and $B$ are the Fourier transforms of the interrogation windows, $\delta x$ and $\delta z$ are pixel shifts in the lateral and axial directions, * is the complex-conjugate, and $\mathcal{F}^{-1}$ is the inverse Fourier transform. Conventionally, the displacement $(\Delta x, \Delta z)$ is given by the location of the single Gaussian shaped peak in the correlation function $r$ (arg max(|r|)); for this to be evident, velocity gradients within an interrogation window must be negligible.

The formulation for WIA assumes one dimensional velocity and we were therefore not interested in obtaining the velocity profile across the vessel; a single interrogation window was used. It was preferred to many smaller windows because it increased computational speed. For comparison with the catheter-based, Doppler method, which identifies peak velocity, the window was sized to exclude regions near to the wall. For all other purposes, the window was sized to span the entire lumen of the vessel.

A distribution of velocities within a window causes the correlation peak to become broadened and skewed. The correlation coefficient, $r$, represents the convolution of the "particle image" (autocorrelation of an image) with the probability density function of the velocity field within the window; the location of the maximum thus represents the modal displacement rather than the mean. Finding the centroid rather than the peak was used to obtain the mean of all possible displacements (see 26]).

Assuming negligible mean displacement of blood in the $z$ direction, the centroid in the $x$ direction was computed by:

$$\Delta x_c = \sum_{\delta x} r(\delta x, 0)\delta x \Big/ \sum_{\delta x} r(\delta x, 0)$$

where for a window size of $N$, the summation ranges from $-N/2$ to $N/2 - 1$ and $\delta y = 0$. However, the correlation plane contains noise and thus integration limits had to be set for the above equation; for large displacements, the broadened peak is shifted towards the edge of the plane, introducing bias. First the particle image diameter ($\sigma$) was measured from the autocorrelation map (24 pixels, as measured using the 1/$e^2$ rule). For non-reversing flows, we assumed that flow at the wall is zero (no-slip condition) and increases towards the centre. Thus the integration limits were defined as $-\sigma/2$ pixels to the maximum displacement ($\Delta x_{max}$) $+\sigma/2$. To determine $\Delta x_{max}$, the modal displacement was found.

Pixel locking is a known problem of centroid calculations but is not an issue if $\sigma$ is >3 pixels, as is the case here. Precision of the centroid calculation was increased by interpolating the correlation map; this was done by zero padding in the frequency domain before inverse transformation.

Interrogation windows were centered on the lateral position of the vessel where the diameter measurements were made. The window width was set to 40 pixels in the first image, and 40 + twice the expected maximum pixel displacement in the second image. (We anticipated velocities to not exceed 1 m/s [27]); this ensured the tails of the broadened peak were captured for large displacements.

*Cross-sectional mean velocity estimation*

The methods described in the previous section provided a measure of the mean velocity in a 2-D, lengthwise slice of the lumen that included the centre line of the vessel. It is an approximation where there is Womersley rather than Poiseuille flow, and hence reverse flow close to the wall when the mean velocity is low. Furthermore, although the maximum velocity in the 2-D slice is the same as the maximum in the 3-D vessel, that is not necessarily true of the mean velocity. Considering the case of Poiseuille flow, for example, the mean is 0.75 times the maximum in a thin 2-D slice, where the velocity profile is a parabola, but 0.5 times the maximum in the 3-D vessel, where the velocity profile is a paraboloid and contains relatively more of the near-wall fluid.

The Poiseuille-based 2-D velocity and Womersley-based 3-D velocity are nearly linearly related (Supplementary Figure 3). The former was used in calculations of wave intensity, where only relative values are required to compare between P-U and D-U methods, and to examine effects of ventricular dysfunction, where only before- vs after-drug comparisons were made. However, absolute numbers was required when calculating wave speeds, to permit comparison with previous values. When undertaking these calculations, the velocity data were further processed to allow for the 3-D, Womersley type flow, as follows:

The inverse Womersley method [28] was applied to the measured flow waveforms after ensemble averaging. Waveforms were decomposed into their Fourier components as,

$$U_{raw}(t) = Re\left(\sum_{j=1}^{m} \hat{a}_j e^{i\omega_j t}\right)$$

where $\hat{a}_j$ and $\omega_j$ are the amplitudes and angular frequencies of the *j*th harmonics. $m = 50$ harmonics were used to reconstruct the waveform. Assuming axisymmetric flow through a rigid cylindrical tube, cross-sectional mean velocity can be computed as

$$U(t) = Re\left(\sum_{j=1}^{m} \hat{a}_j \, G(\alpha_j) \, e^{i\omega_j t}\right)$$

where

$$\alpha_j = \bar{R}\sqrt{\frac{\omega_j}{\nu}}$$

is the Womersley number, $\bar{R}$ is the mean radius and $\nu$ the kinematic viscosity of the blood.

To obtain the volumetric mean flow from the Doppler data (centreline velocity):

$$G(\alpha_j) = \frac{1}{1 - J_0(i^{3/2}\alpha_j)}\left(1 - \frac{2J_1(i^{3/2}\alpha_j)}{i^{3/2}\alpha_j J_0(i^{3/2}\alpha_j)}\right)$$

And from the velocimetry data (line average velocity):

$$G(\alpha_j) = \frac{1}{\int_0^{\bar{R}} 1 - \frac{J_0(i^{3/2}\alpha_j\ r/\bar{R})}{J_0(i^{3/2}\alpha_j)} dr}\left(1 - \frac{2J_1(i^{3/2}\alpha_j)}{i^{3/2}\alpha_j J_0(i^{3/2}\alpha_j)}\right)$$

where $J_0$ and $J_1$ are zeroth and first order Bessel functions. For Poiseuille flow, these equations simplify to scaling factors of 0.5 and 0.75, respectively, as required.

*Wave intensity analysis*

Waveforms were smoothed using a 2nd order Savitsky-Golay filter with window length of 19 ms. Successive beats in the pressure, diameter and velocity waveforms were ensemble averaged using the ECG R wave as a fiducial marker; only complete cardiac cycles were included.

Net wave intensities were calculated as

$$dI = \frac{dP}{dt}\frac{dU}{dt}$$

$$_{n}dI = \frac{dD}{dt}\frac{dU}{dt}$$

the prefix $n$ refers to the non-invasive diameter formulation and $dt$ is the time between two consecutive samples. $dI$ was normalised by $dt^2$ to make the magnitude of $dI$ independent of the sampling frequency.

The peak magnitudes, area under the curve ("wave energy") and timings with respect to the R wave of the ECG were calculated for the three dominant waves (systolic, reflected and diastolic). The start and end of the waves were identified by the zero crossing of the wave intensity.

Wave speed $c$ was calculated using single point "loop methods" [29]. In plots of ensemble-averaged $P$ (or $\ln(D)$) against $U$, there is a linear portion of the loop in early systole whose gradient can be used to determine $c$ according to:

$$c_{PU} = \frac{dP}{\rho dU}$$

$$c_{\ln(D)U} = \frac{1}{2}\frac{dU}{d ln(D)}$$

where $\rho$ is the density of the blood (1044 kg/m$^3$). The gradients during the first 10 ms (rabbit) or 50 ms (human subject) of ejection were calculated through linear fitting. The equations are based on the assumption of a reflection free period in early systole. A third wave speed [30]:

$$c_{\ln(\mathrm{D})P} = \sqrt{\frac{1}{2\rho}\frac{0.5dP}{dln(D)}}$$

was also calculated when $P$ and $D$ were recorded simultaneously. Not only is this value free of potential velocity errors, but the equation's derivation does not assume unidirectional wave travel.

***Human study***

Values of D and U were obtained by non-invasive ultrasound of the carotid artery of a healthy young subject, using the parameters described above except that MI was increased to 0.4. Ethics approval 18/LO/1724 was obtained after the study had been reviewed by three independent academic referees and an NHS Research Ethics Committee (ArterioWave - Arterial Pulse Waves in Heart Failure, IRAS Project ID 248724). Informed consent was given.

***Statistics***

P values were obtained by Student's paired t-test unless otherwise stated. Bland Altman plots were created in R (The R Project for Statistical Computing).

## Results

Figures 1a and 1b show, respectively, invasive pressure measurements and non-invasive diameter measurements for each animal. The two types of waveform clearly have the same shape, but the relation between them is not expected to be perfectly linear: the elasticity of the vessel decreases as diameter increases ("strain stiffening"), and viscous effects cause a lag between changes in pressure and changes in diameter. The non-linearity and hysteresis are visible when D was plotted against (Figure 1c); data for D were filtered as described above.

a)

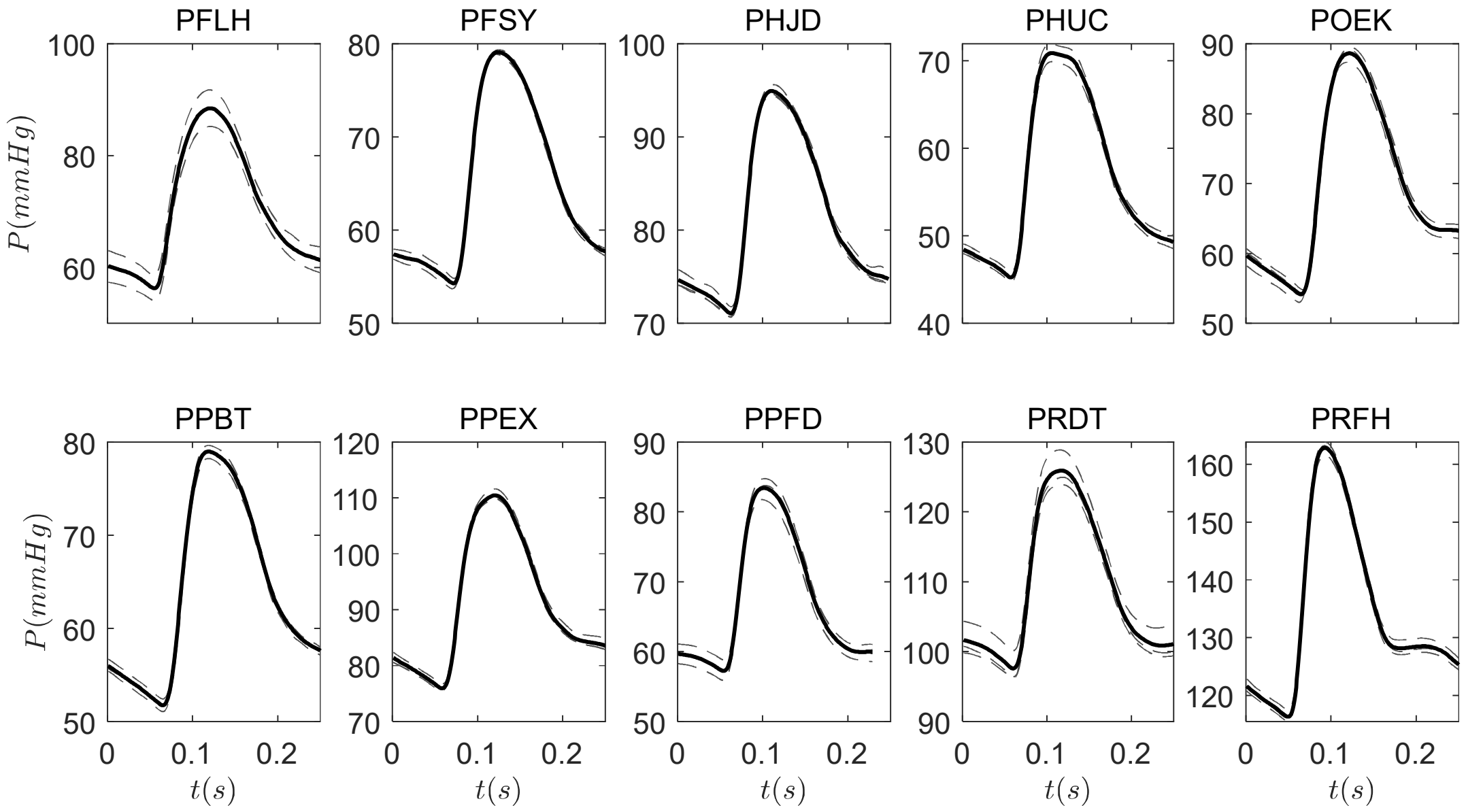

b)

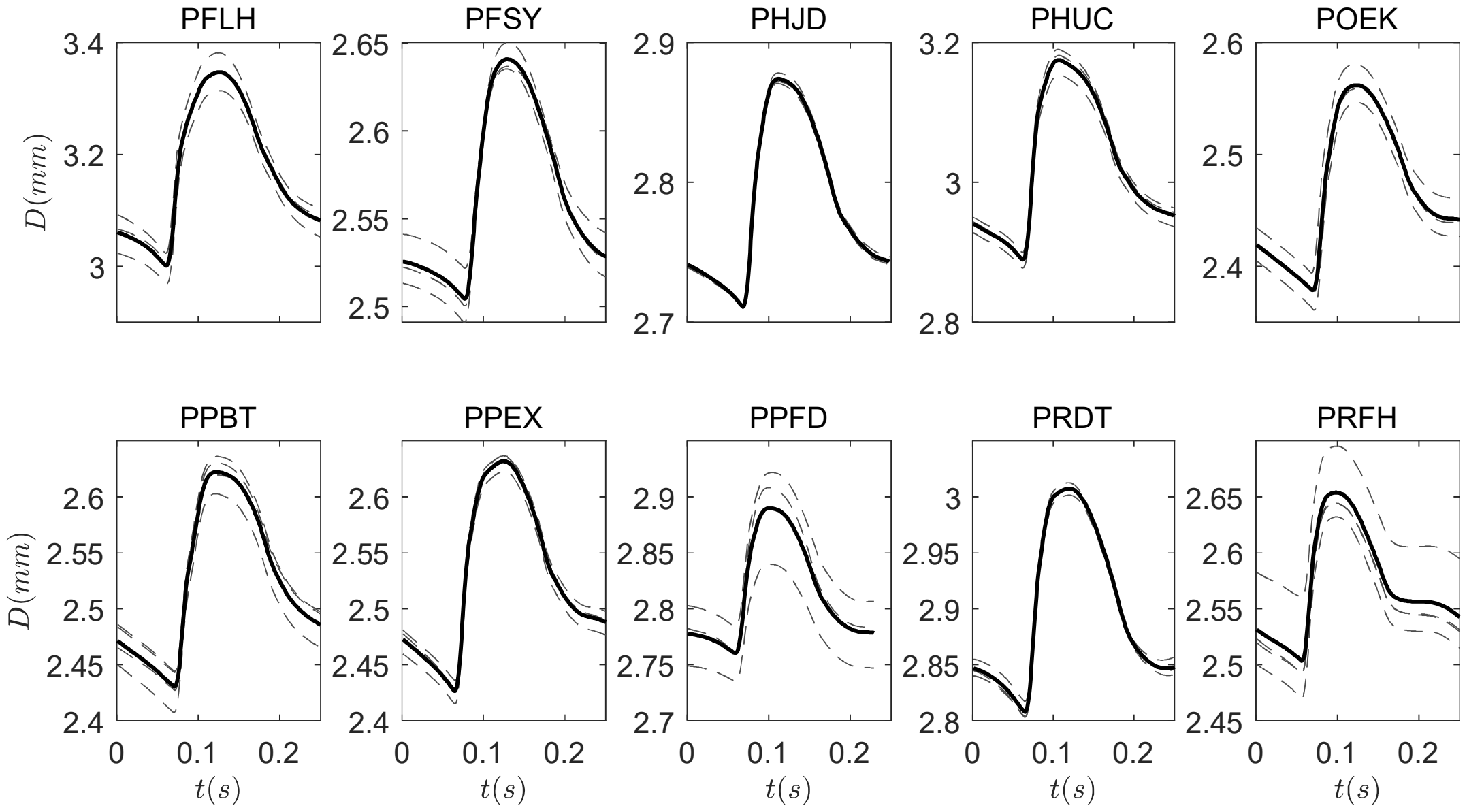

c)

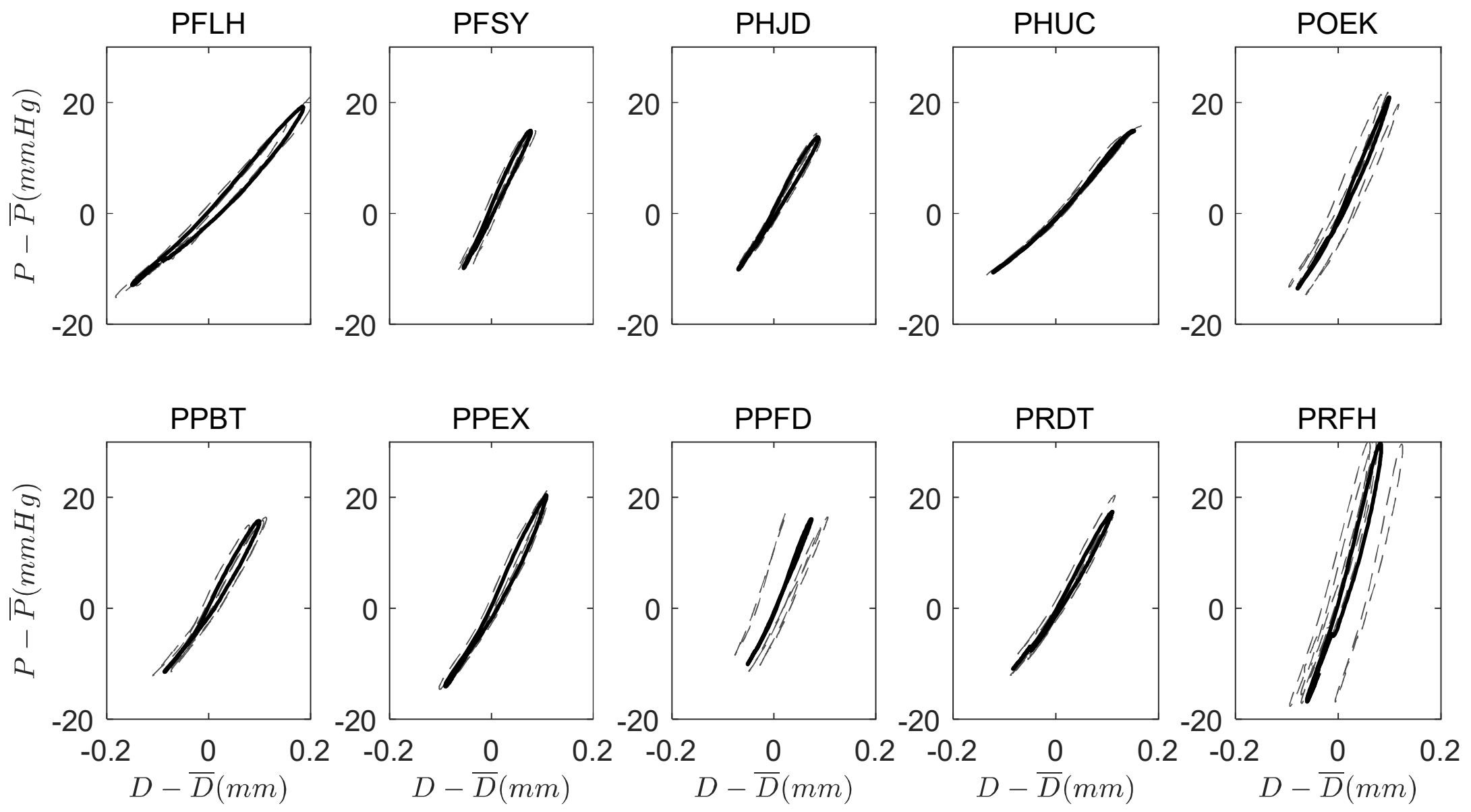

Figure 1. (a) Pressure and (b) diameter measurements, and (c) the relation between them. The four-letter codes identify individual rabbits. Dashed lines represent the ensemble average of several heartbeats for each repeat, and the solid line is the average of all (n=3-4) repeats. Note the different y-axis scales between animals in (a) and (b). In (c), mean pressure and mean diameter have been subtracted from instantaneous values. Time during the cardiac cycle runs clockwise around each loop. Rabbit PRFH has high pressure, anomalous pressure and diameter waveforms, and a stiff vessel.

Figure 2 shows (a) invasive velocity measurements and non-invasive velocity measurements, and (b) the relation between them. The two methods were time-aligned by the R wave of the ECG. The region of interest in the B-mode images of the non-invasive system was reduced to a small box at the centre of the vessel, to permit a fairer comparison with the invasive Doppler system. (Note that neither should be used in calculations of wave speed and wave intensity; a correction is required to give the mean rather than maximum velocity – see above and Mynard et al., 2018.) In (b), the non-invasive data have been filtered as described above.

It is apparent in (a) that the invasive and non-invasive velocity measurements, whilst in broad agreement, show a systematic difference: the invasive curves resemble smoothed versions of the non-invasive ones. The sampling frequency of the Volcano Combowire system and the additional low pass filtering mean that the system cannot capture the fast early-systolic accelerations in the rabbit. There additionally appears to be a scaling error in rabbits PPEX, PPFD, PRDT and PRFH, which was most likely caused by the difficulty of perfectly aligning the orientation of the catheter with the mean flow direction. (The Doppler probe at its tip obtains velocity information only along the beam axis)/ Because of this issue, B-mode derived velocity data are used in all subsequent analyses.

a)

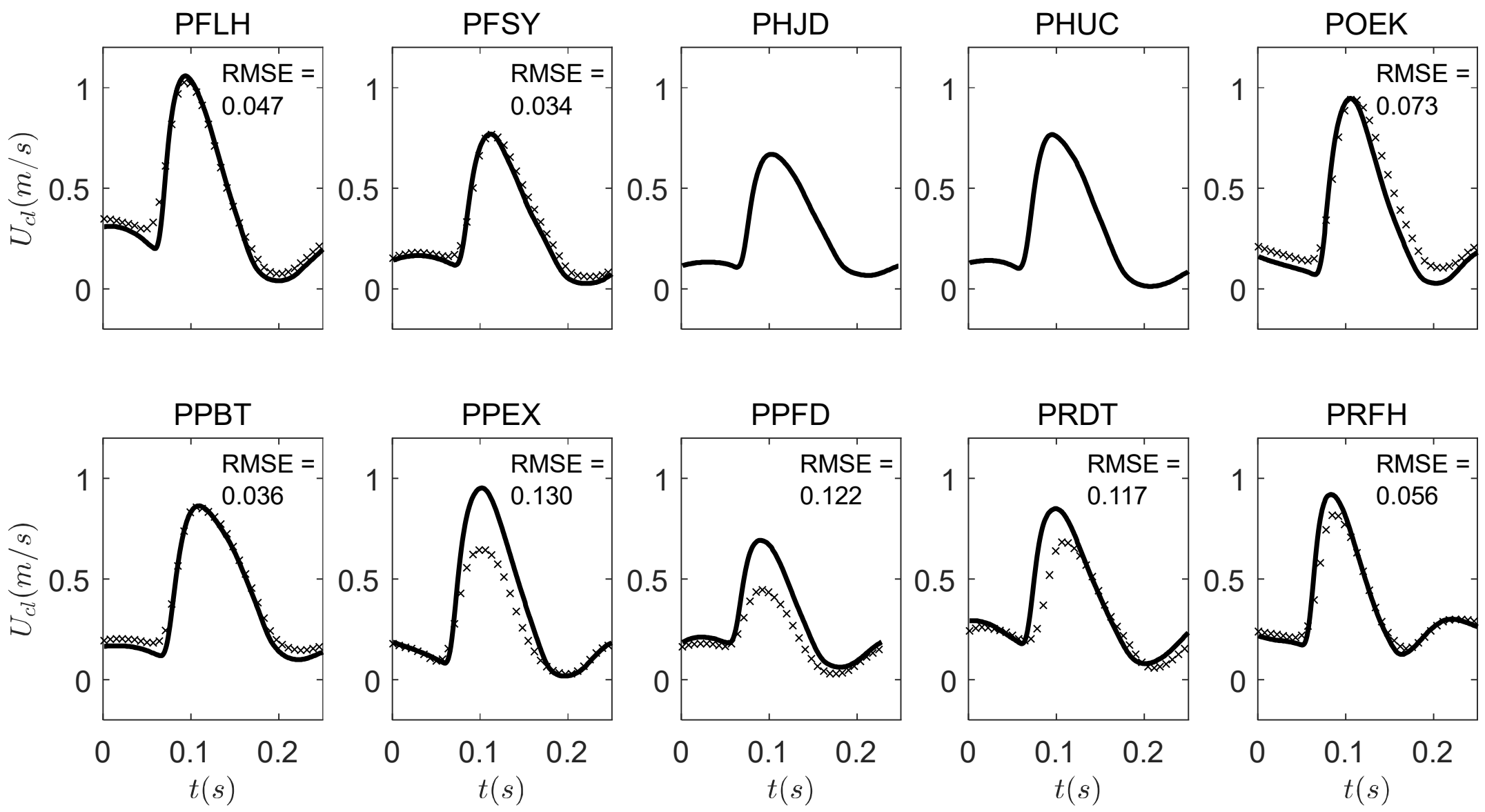

b)

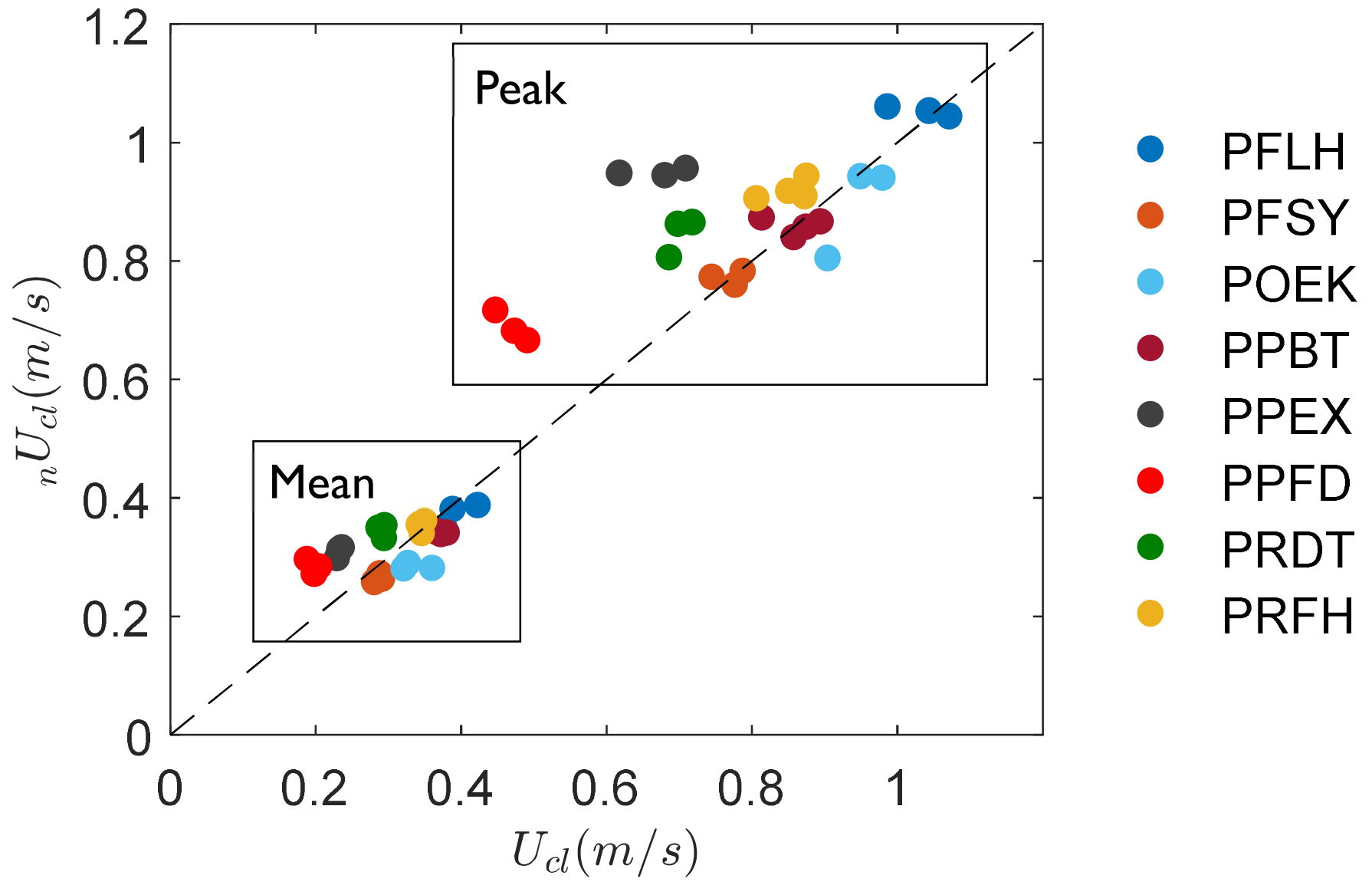

Figure 2. (a) Catheter-based Doppler (crosses) and non-invasive B-mode (solid) measurements of blood velocity, and (b) the relation between the mean and peak velocities measured by the two methods (the dashed line being the line of identity). The four-letter codes identify individual rabbits. (Catheter-based Doppler velocity traces were not obtained in two of the rabbits shown in Figure 1). In (a), dashed lines represent the ensemble average for each repeat and the solid line is the average of all (n=3-4) repeats; the root mean square error (RMSE) obtained when comparing the methods are shown for each rabbit.

Figure 3 shows calculations of wave speed, c, by three different “loop” methods – the PU loop, ln(D)U loop and ln(D)P loop – for each rabbit, and the relation between the values obtained by the different methods. D and U, obtained from the non-invasive ultrasound methods, were filtered as described above. The wave speed given for each rabbit is the mean of the wave speed obtained in each repeat, not the wave speed computed from the mean loop.

Compared to the ln(D)P method, which is considered to gold standard (see above), the PU method consistently underestimated wave speed and the ln(D)U method consistently overestimated it. However, the mean errors were small: -0.55 and +0.36 m/s, respectively, when compared to a best estimate of 5.02 m/s (the mean value obtained with the ln(D)P method). Wave speeds were calculated for the first 10ms of ejection, where reflections are expected to be lowest. Nevertheless, the systematic elevation of PU-derived wave speeds and the systematic lowering of ln(D)U-derived wave speeds, compared to those obtained by the ln(D)P method, are what would be expected if errors resulted from reflections. Of course, inaccuracies in D and P measurement or processing may also be responsible. The errors do not appear to scale with wave speed but remain relatively constant for all values obtained with the ln(D)P method.

a)

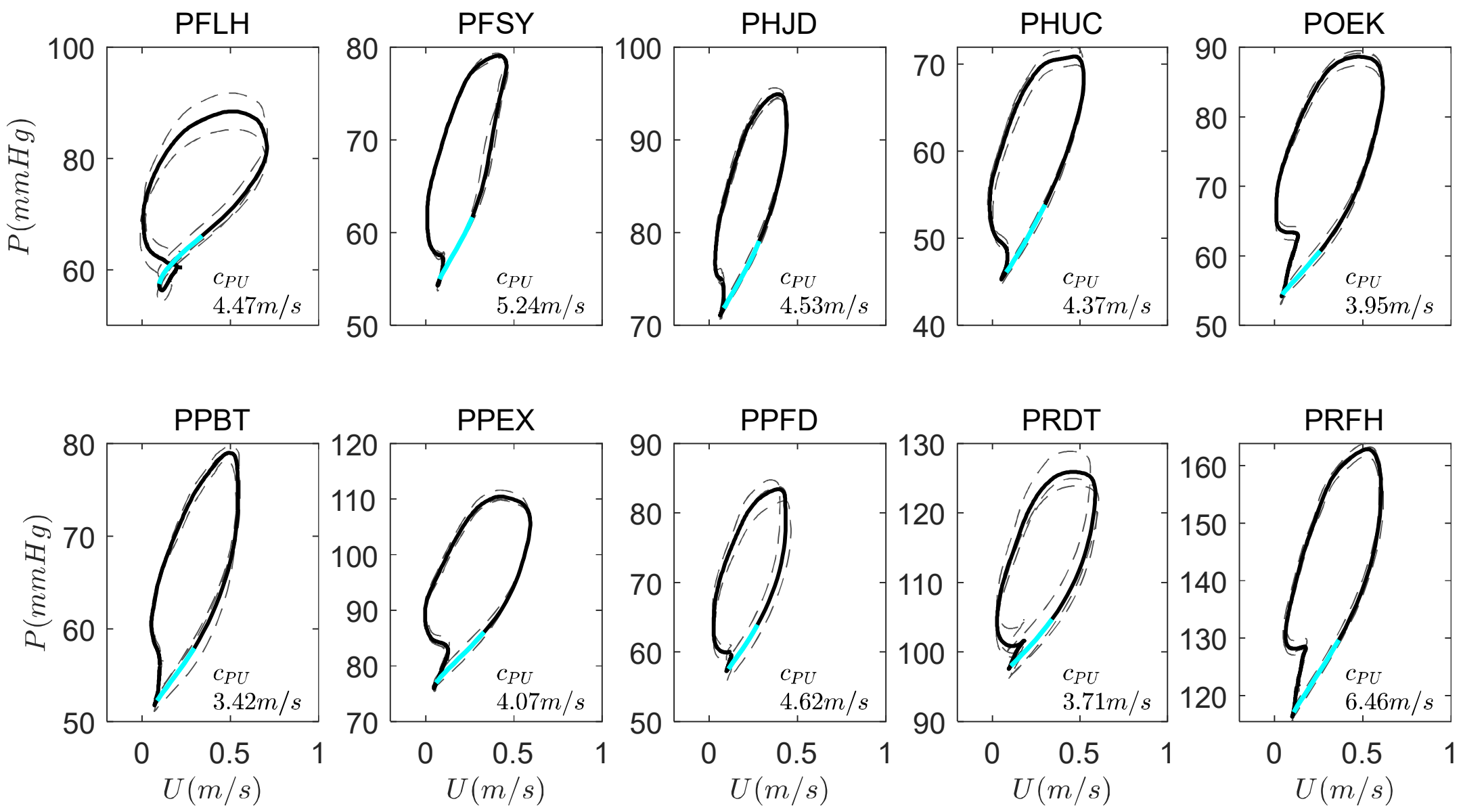

b)

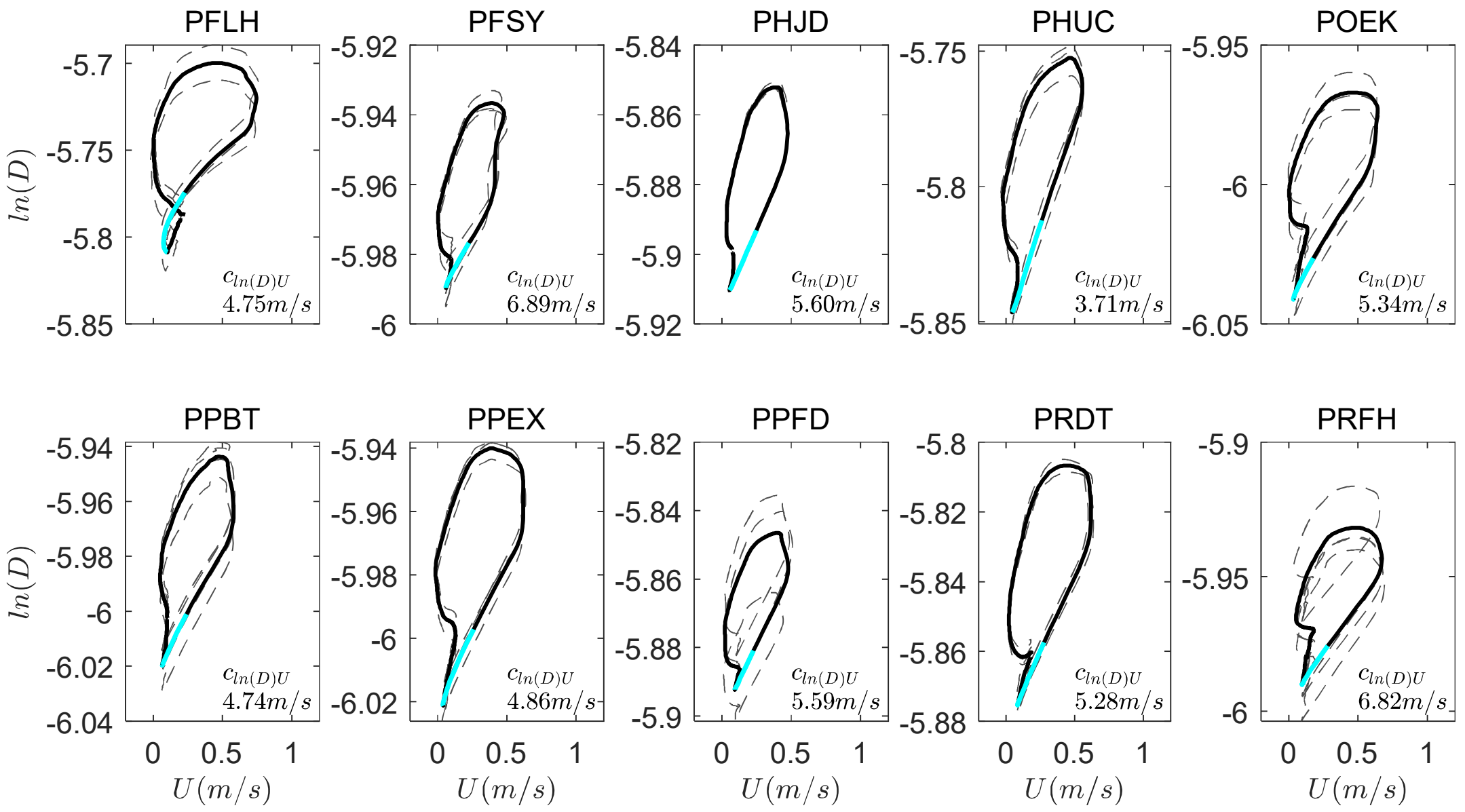

c)

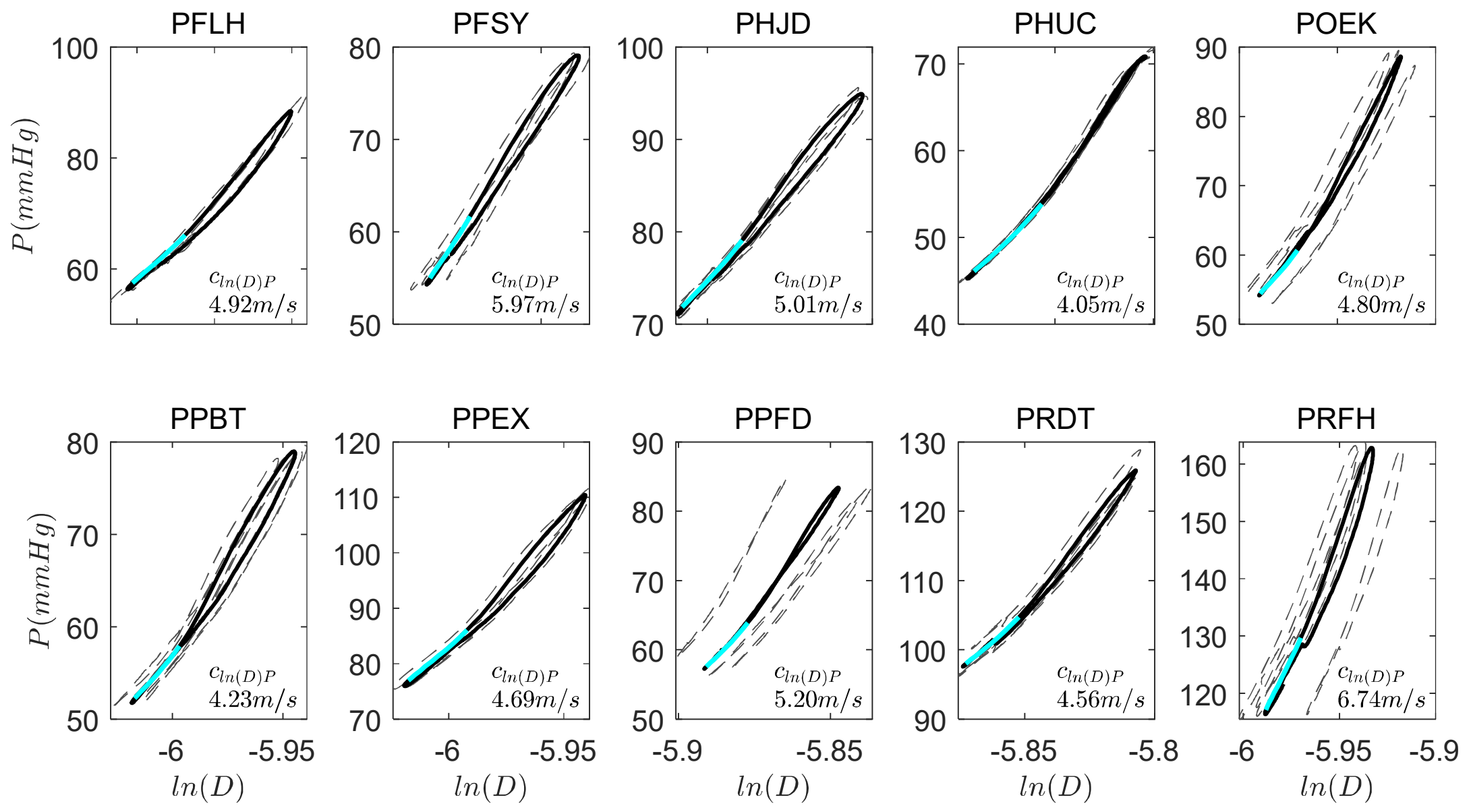

d)

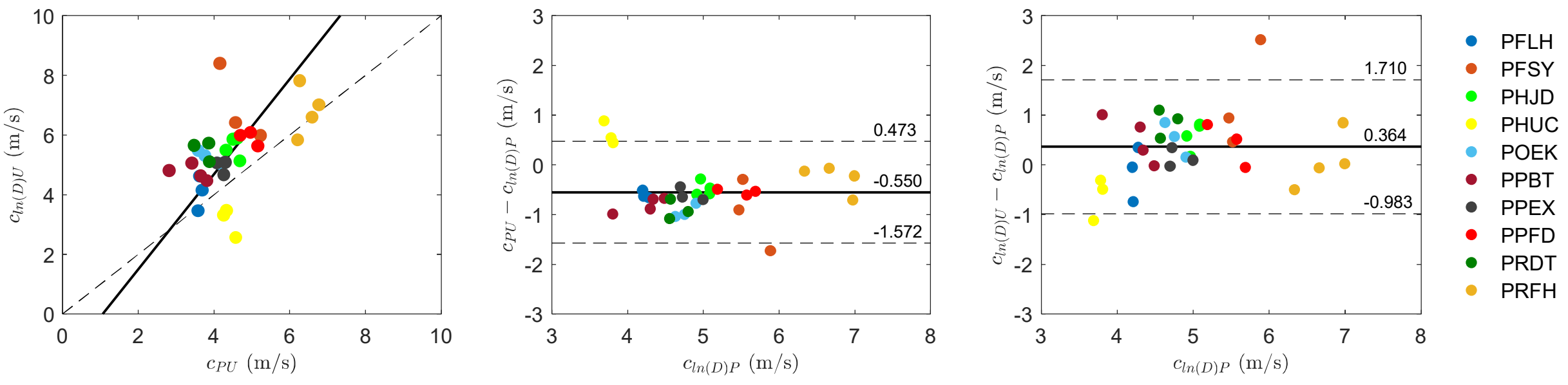

Figure 3. (a) PU loops, (b) ln(D)U loops and (c) ln(D)P loops for each rabbit, and (d) comparisons between the methods. The four-letter codes identify individual rabbits. In (a-c), dashed lines represent the ensemble average for each repeat and the solid line is the average of all (n=3-4) repeats. Wave speed, c, for each rabbit was calculated over the first 10ms of ejection, shown in blue. Note the different y-axis scales between rabbits. All methods again show anomalous results for rabbit PRFH. In (d), the left-hand plot shows orthogonal ("Deming") regression of mean wave speeds obtained for each rabbit by the PU and ln(D)U methods (the dotted line showing the line of identity); the centre and right-hand plots assume the wave speed obtained by the ln(D)P method is correct and show, respectively, discrepancies with the PU and ln(D)U methods, using Bland Altman plots with repeated measures.

Figure 4 shows wave intensities calculated by both the P-U and D-U methods, and a comparison between the two methods. WI values were divided by $dt^2$ in order to make them independent of the sampling interval and hence directly comparable with other studies. D and U were filtered as described above. Note that these intensities are the sum of forward- and backward travelling wavelets (with positive and negative intensities, respectively) at each time point. Three waves are evident during the cardiac cycle. There is a large forward-travelling wave (“W1”) followed immediately by a smaller backward-travelling reflected wave (“R”), and then a further forward-travelling wave (“W2”) before a return to baseline. The backward travelling wave comprises reflections from distal sites. When comparing the two methods, there was a clear correlation between the peak intensity and the area under the intensity curve (the “wave energy”) for each wave.

a)

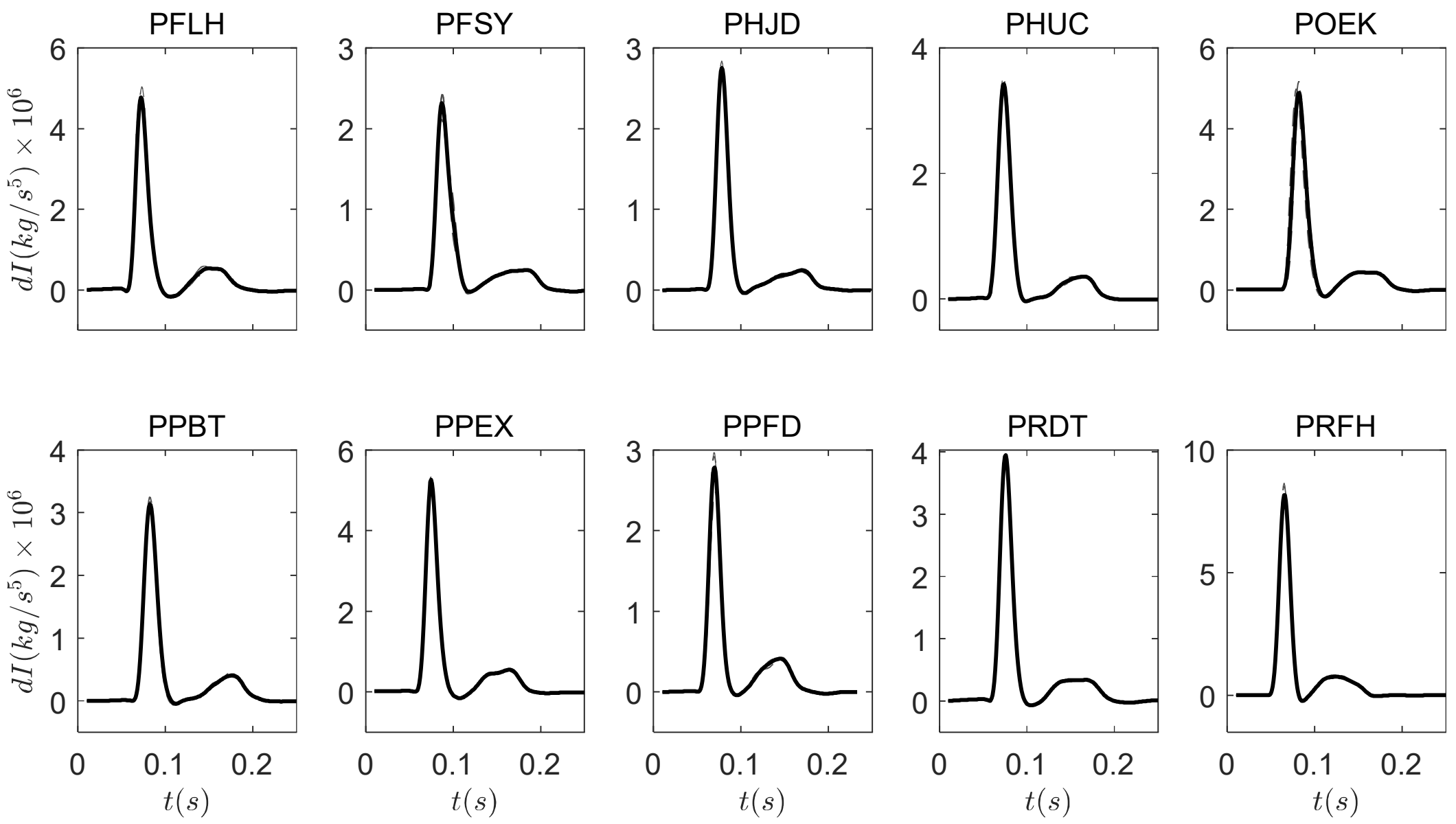

b)

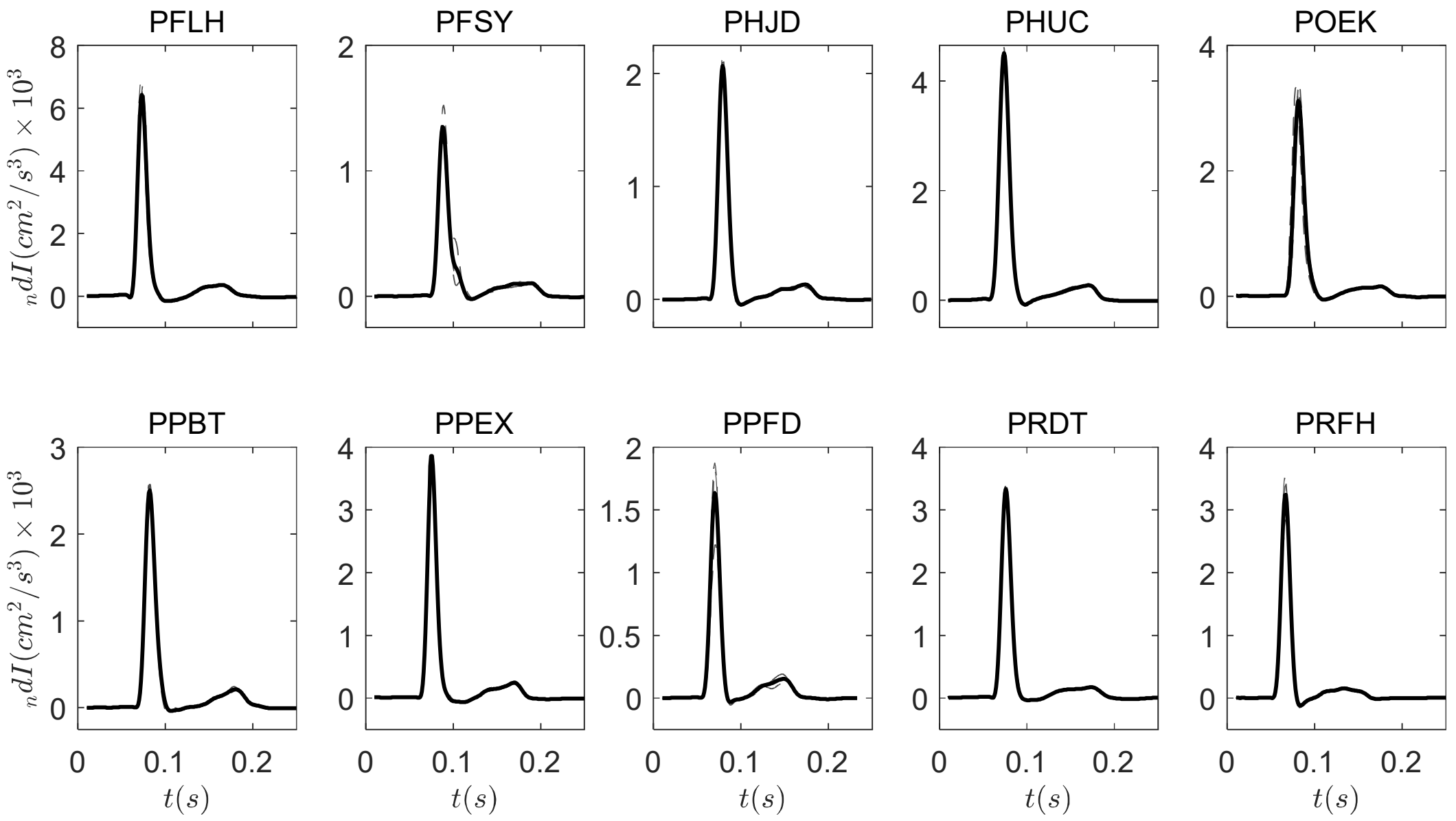

c)

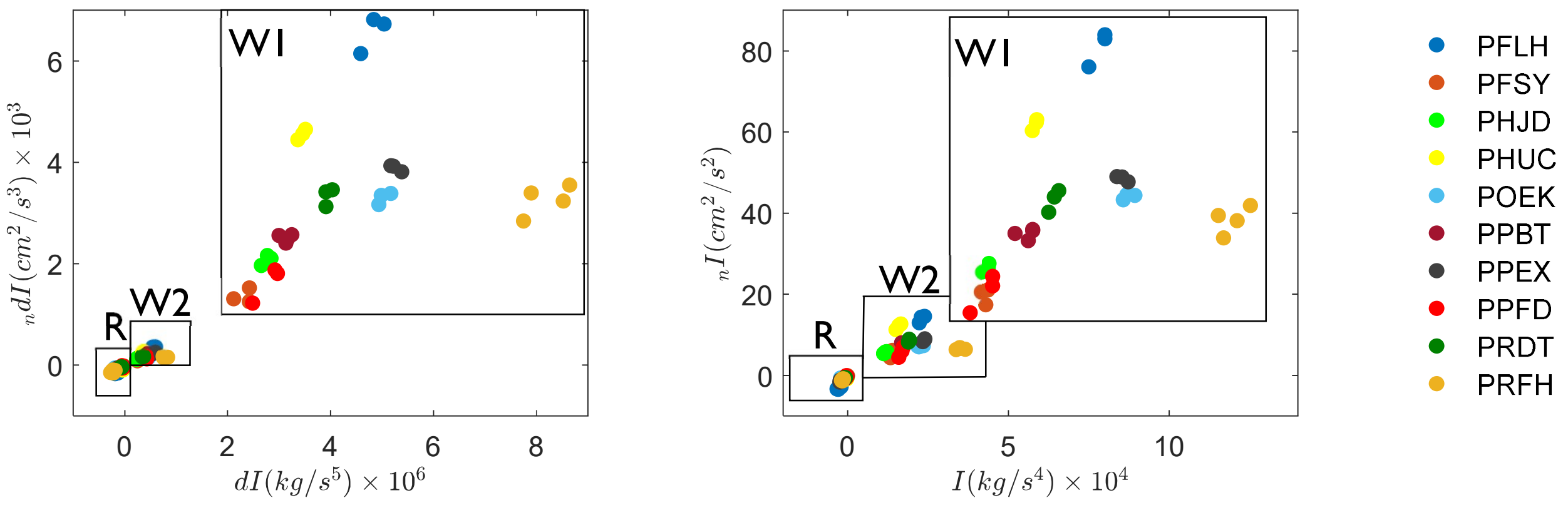

Figure 4. Wave intensities derived from (a) pressure and velocity data and (b) from diameter and velocity data, and (c) comparisons between the two methods for (left) the peak intensity and (right) the energy of each wave. The four-letter codes identify individual rabbits. In (a) and (b), dashed lines represent the ensemble average for each repeat and the solid line is the average of all (n=3-4) repeats. Three waves are visible in each case: a forward-travelling “W1” wave, followed immediately by a small reflected “R” wave, and then a further forward-travelling “W2” wave. In (c), the wave intensities (left) and energies (right) derived from pressure and velocity are on the x-axes and those derived from diameter and velocity data are on the corresponding y-axes. T=0 corresponds to the R wave of the ECG (not to be confused with the R wave of the pulse). There is a clear positive relation in both plots. Data for rabbit PRFH appear to lie on a different line from the remaining animals, reflecting its stiffer artery.

Figure 5 shows the effect on wave intensity of esmolol, a cardio-selective, $\beta_1$-adrenergic blocker that reduces the force and rate of cardiac contraction, assessed by the invasive and non-invasive methods. Figure 5(a) gives illustrative values of wave intensity in rabbit PFLH during the procedure. Transient dips in wave intensity are evident after each esmolol bolus, and they increase in size with increasing dose. A delay in the arrival of the W1, R and W2 waves is also evident after the administration of each bolus, with the largest delay occurring at the highest dose.

Figure 5(b) shows the peak intensity of the W1 and W2 waves, calculated by both the pressure-velocity and diameter-velocity methods in 9 rabbits. U and D data were filtered. Figure 5(c) shows the areas ("wave energies") under the W1 and W2 waves. Rabbit PPBT was not included in the analysis as it required an anesthetic top up during the period of esmolol administration, rendering the results unreliable. Data are presented as mean±SD to show variability within the sample. Diameter-velocity data for two control rabbits administered vehicle rather than esmolol are also shown.

The dips in wave intensity are again evident after each esmolol bolus for both waves, both wave parameters and both methods. No such trends are seen in the control animals. Statistical tests were performed by comparing the first measurement after each dose of esmolol with the preceding measurement. For the experimental group, the drop in W1 was significant ($p<0.05$) at all doses, for peak intensity and wave energy, with both methods. The drop in W2 wave energy was also significant in all these cases. For W2 peak intensity, the effect was not significant ($p>0.05$) for the lowest esmolol dose assessed with the pressure-velocity method and for all doses with diameter-velocity method. The fact that statistical significance was obtained for wave energies but less consistently with peak intensities is likely due to alignment issues – see Discussion

Variation in the time of arrival of the W1 wave are shown in Figure 5(d). Data for rabbits PPEX and PRDT at t=25 mins were omitted due to a poor ECG signal. A clear delay is seen after each bolus, and the effect was significant ($p<0.05$) at all doses. Delays of similar magnitude, again all significant, were seen in the interval between the W1 and W2 waves.

a)

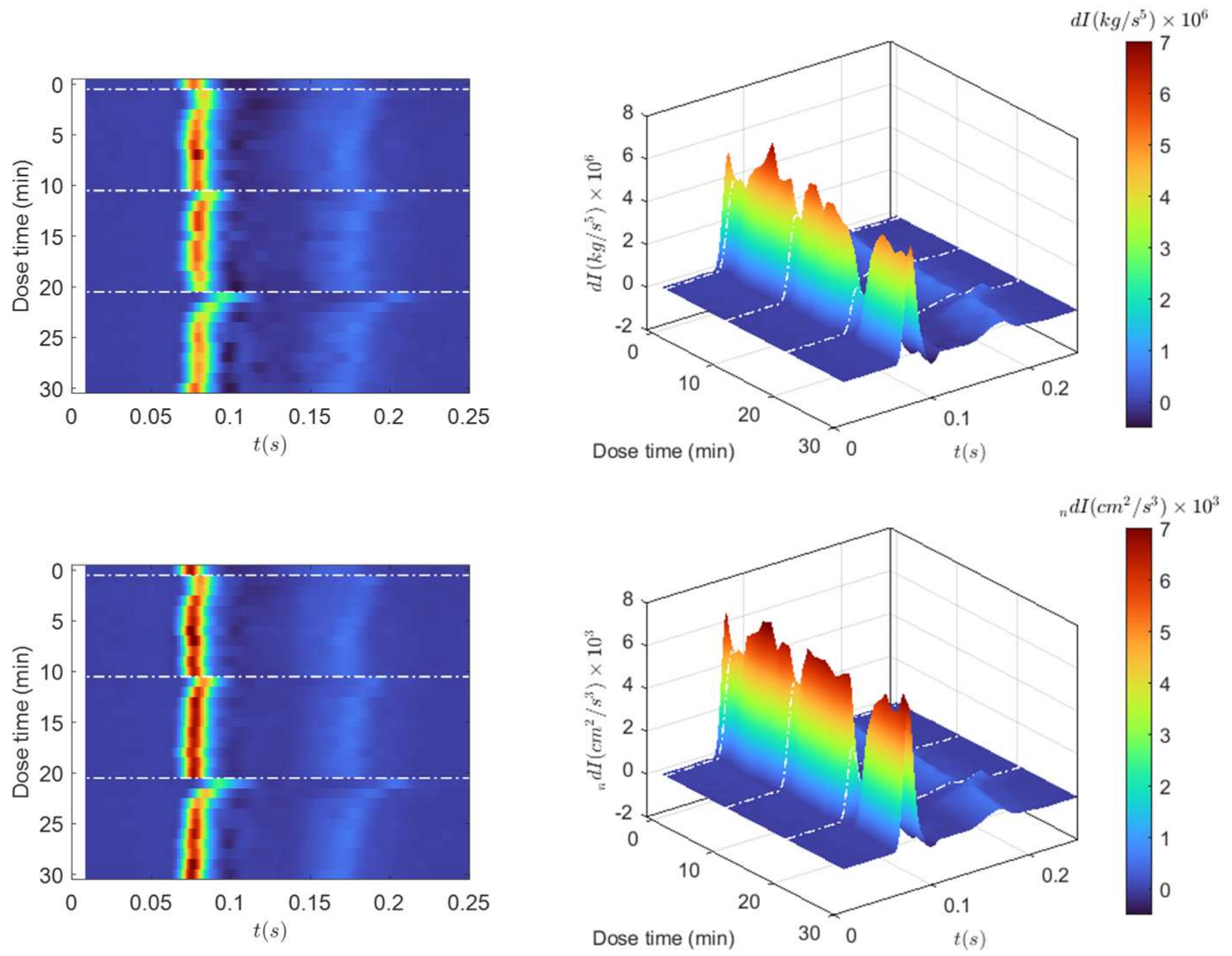

b)

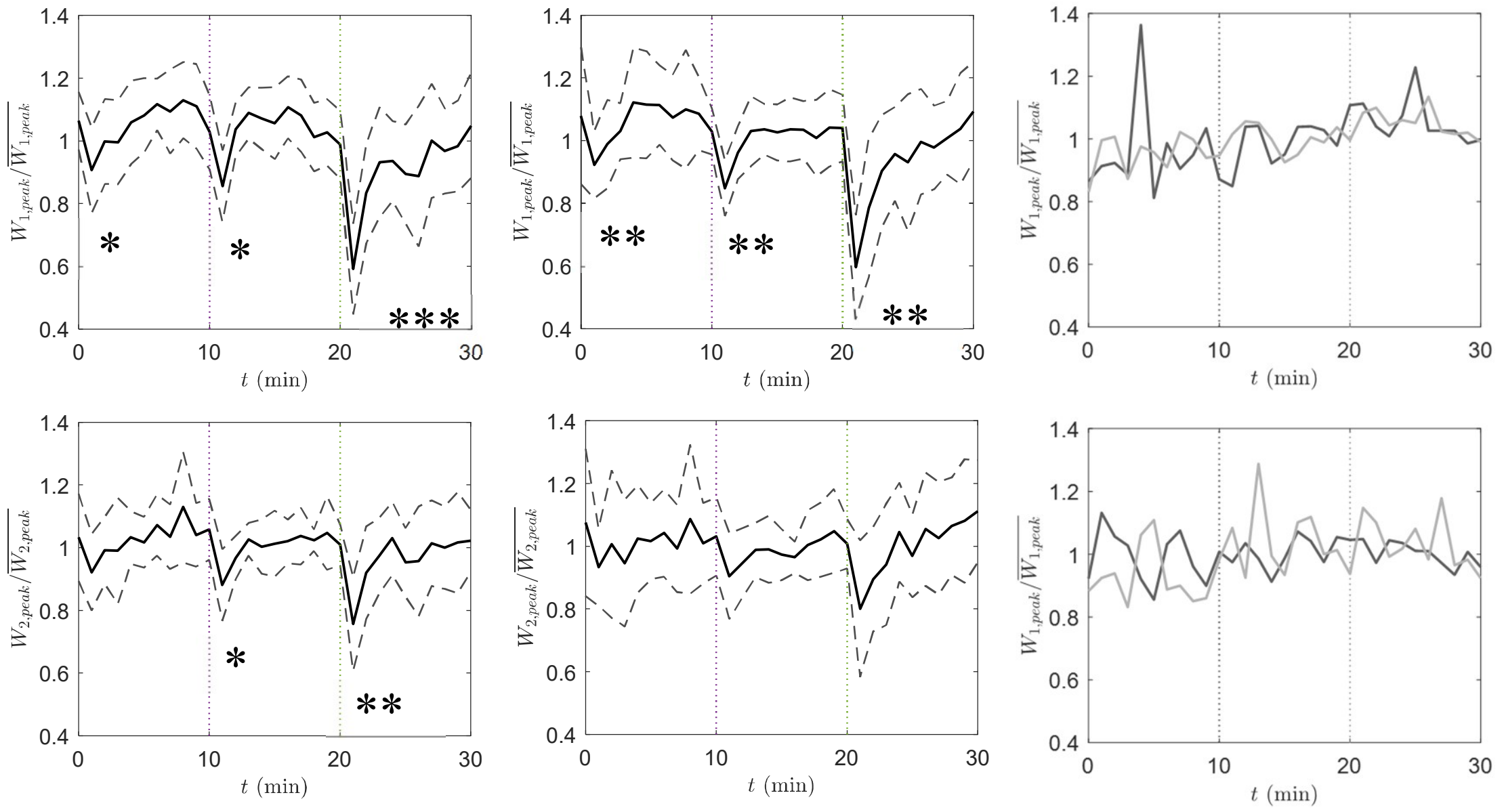

$W_{1,peak}/\overline{W_{1,peak}}$
$W_{2,peak}/\overline{W_{2,peak}}$
$t$ (min)

c)

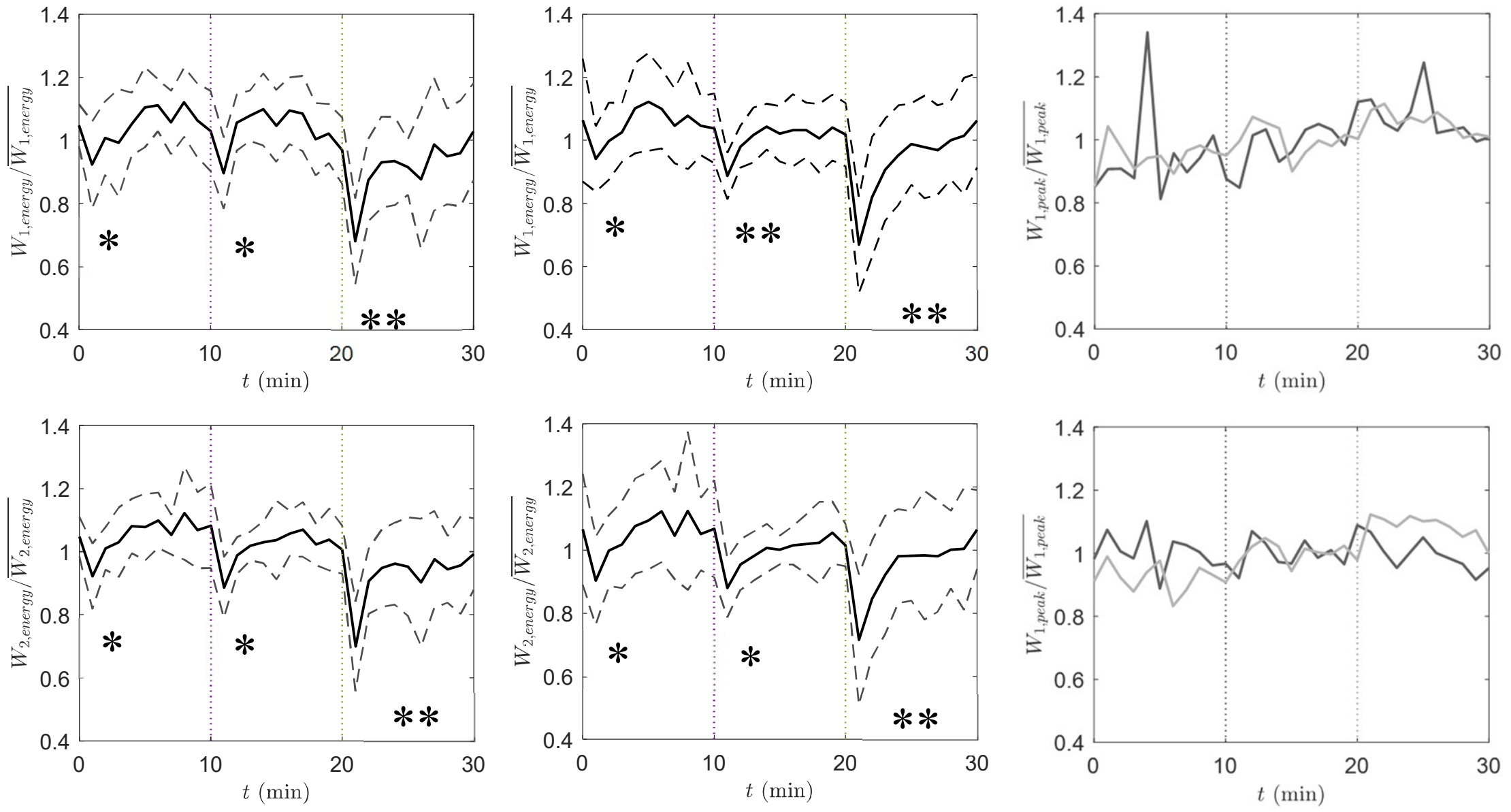

$W_{1,energy}/\overline{W_{1,energy}}$
$W_{2,energy}/\overline{W_{2,energy}}$
$W_{1,peak}/\overline{W_{1,peak}}$
$t$ (min)

d) timing

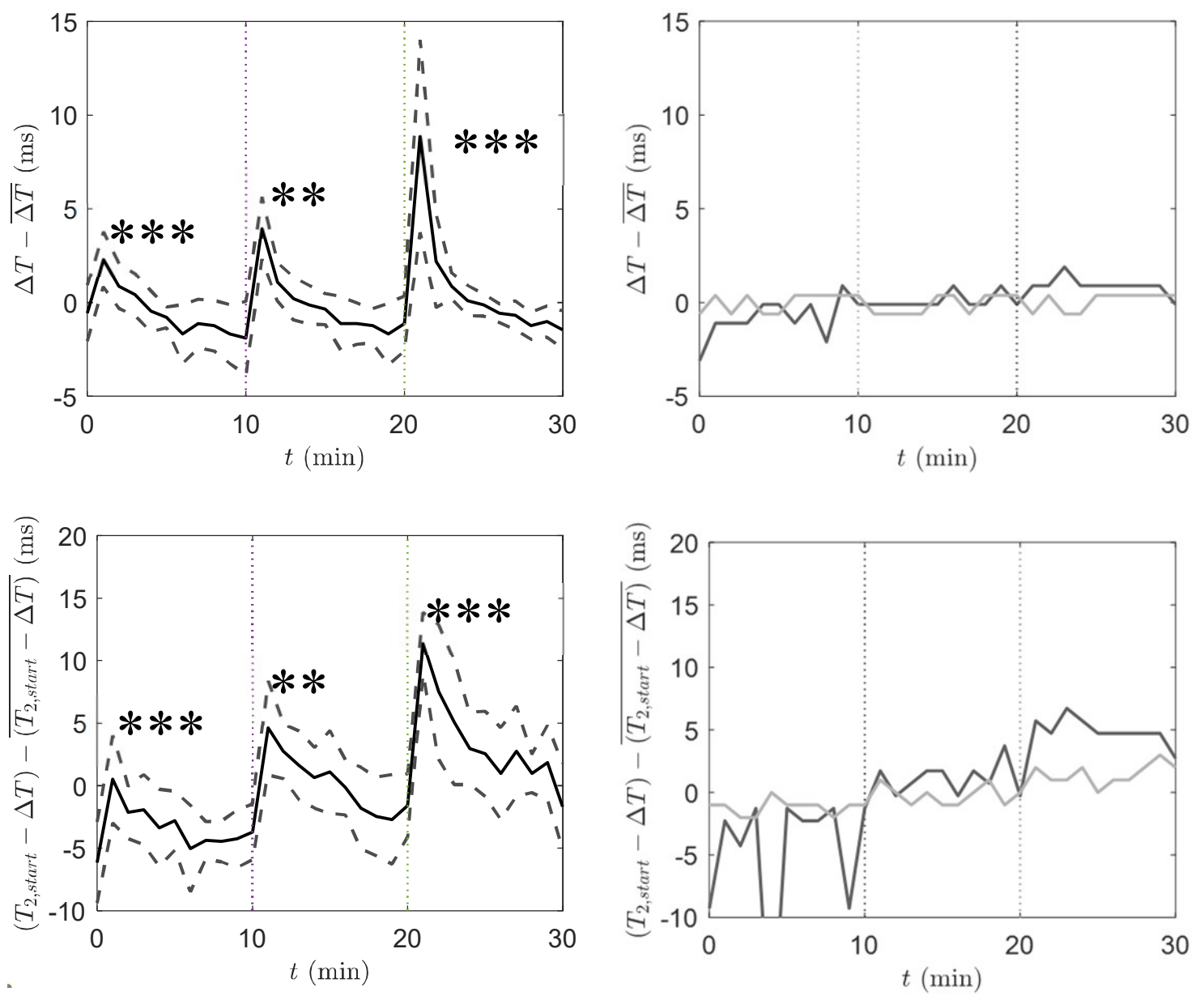

Figure 5. (a) Values of wave intensity in a single rabbit during administration of esmolol, determined by the invasive pressure-velocity method (top) and the non-invasive diameter-velocity method (bottom). The two horizontal axes in the left-hand panel both represent time, but at different scales – one indicates a single cardiac cycle, with alignment at t=0 by the R-wave of the ECG, and the other represents the 30-minute duration of the experiment, with esmolol administered at increasing doses immediately after recordings were made at 0, 10 and 20 minutes. Wave intensity is represented by the height on the vertical axis and is also colour coded. The right-hand panel shows the same data presentation but viewed from the top. (b) The peak intensity and (c) the wave energy of the W1 and W2 waves in rabbits administered esmolol, determined by the invasive pressure-velocity method (left) and the non-invasive diameter-velocity method (centre). Both left and centre plots show mean (solid line) ± SD (dashed lines) for n=9 rabbits. The right-hand plots show individual data for two rabbits (RNLX, RNNE) administered vehicle only, determined by the non-invasive diameter-velocity method. (d) Variation in the time of arrival of the W1 wave, relative to the ECG R-wave (top left) and in the time interval between the arrival of the W1 and W2 waves (bottom left). Data (mean±SD, n=7-9) are normalised by the mean arrival time throughout the 30-minute period in the same rabbit. The same variables are also shown for the two control rabbits (top right and bottom right).

Figure 6 demonstrates the feasibility of using the non-invasive method in the carotid artery of human subject subjects.

a)

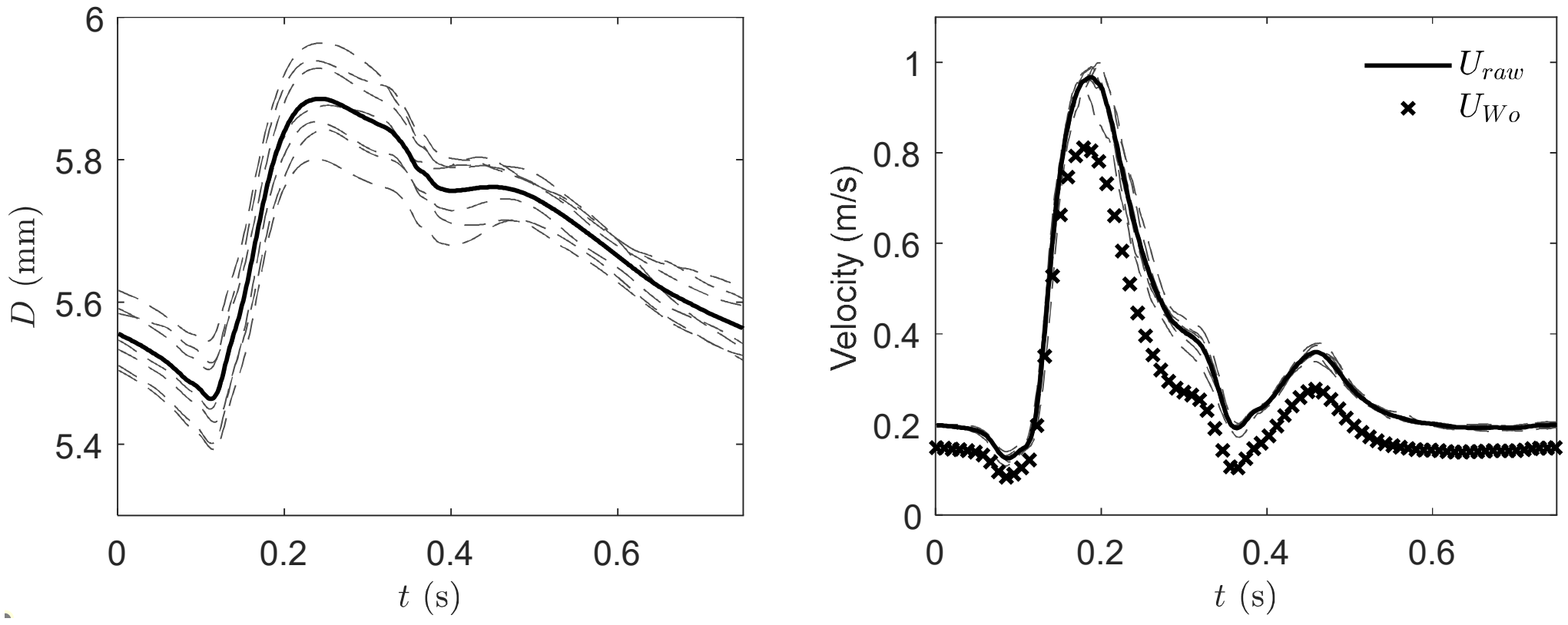

b)

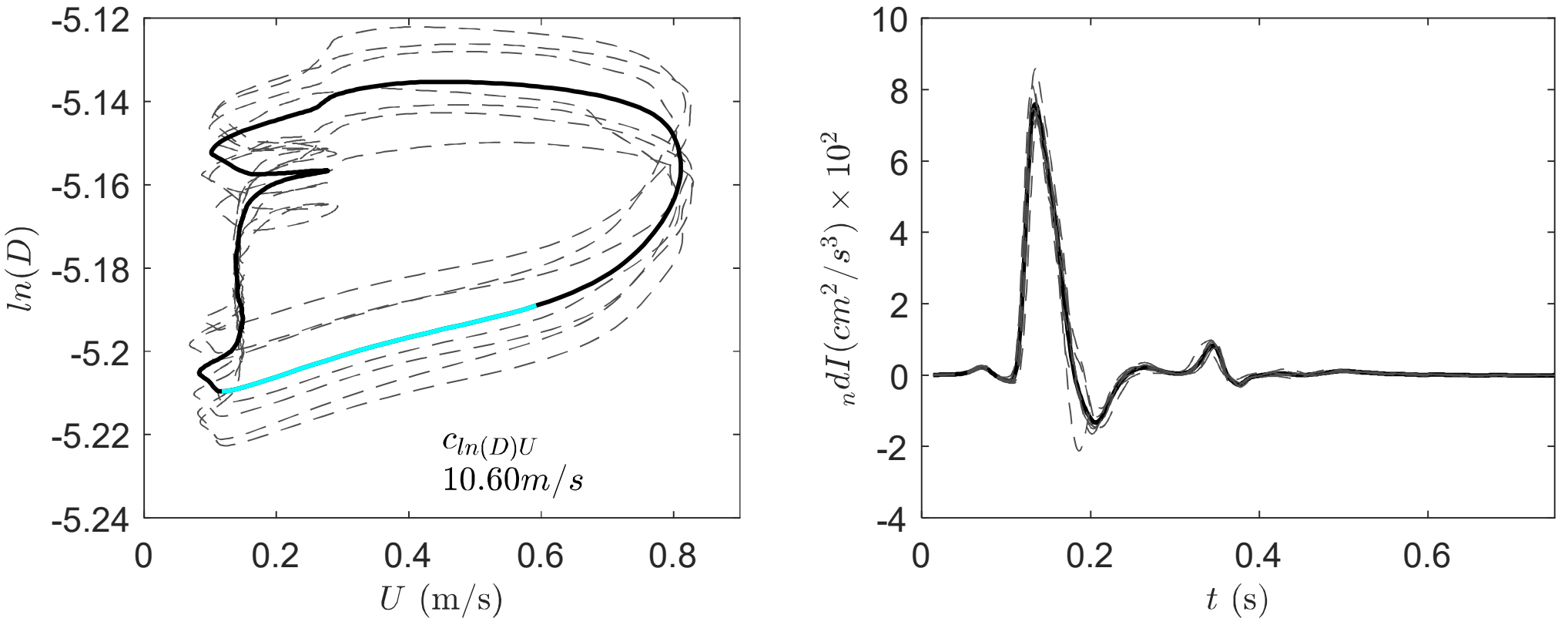

Figure 6. (a) Carotid artery diameter and (b) blood velocity measured in a healthy human subject. Both the raw velocities and velocities computed according to the 3-D Womersley formulation are shown. Data were aligned and ensemble averaged using the R-wave of the ECG as a datum. (c) The corresponding ln(D) vs U loop. The wave speed was computed from the slope of the first 50ms of ejection, shown in blue. (d) plot of ensemble averaged net wave intensity through the cardiac cycle. In all panels, dashed lines represent the ensemble average for each repeat and the solid line is the average of all repeats.

## Discussion

An advantage of calculating wave intensity from arterial diameter and blood velocity is that both can be measured using non-invasive methods. We developed methods based on ultrasound because its low cost and real-time imaging capability increase clinical utility. Wave intensities are computed from derivatives, which increases the requirement for measurement accuracy and precision. We avoided the use of Doppler ultrasound because of its fundamental limitations in quantifying blood velocity and because accuracy can be further compromised by the difficulty of employing the optimal, orthogonal beam angles for the Doppler and diameter measurements. Instead, changes in diameter and velocity were obtained from the same, successive B-mode images by tracking the movement of scatterers in the wall and blood. Obtaining both measures from the same images precludes the alignment issues apparent when using the current “gold-standard” invasive system, which relies on separate pressure and (Doppler) flow transducers, even though they are mounted on the same catheter.

The new method presents several technical challenges. Blood is a poor ultrasound scatterer at the low frequencies required to penetrate tissue to the depth of conduit arteries. This necessitated the use of singular value decomposition (SVD) to obtain separation of the blood signal from the tissue signal and noise. SVD is computationally intensive, and the manual selection of ranks was slow. Randomised SVD is faster [25], and automated methods for selecting ranks are being developed [31]. A further technical challenge is capturing the fast-moving arterial blood with sufficient temporal resolution. This was achieved by using an ultrafast, plane-wave scanner capable of frame repetition rates of ≈10 kHz. In such systems, the beam is not focused, which accelerates scanning but degrades the image. Image quality is recovered by averaging frames and by interleaving scans at different angles. The net frame rate still exceeds that of conventional, focused scanners.

Velocity estimation is also complex. The use of B-mode imaging means that scattering from red blood cells can theoretically be tracked in two dimensions. In conventional Ultrasound Imaging Velocimetry (UIV), a small field of scatterers is identified within the vessel in one image and then a search is made for the field in a succeeding image that has the best-fit pattern of scatterers; the displacement of the field, divided by the time interval between the two scans, gives the velocity for that region of blood. Repeating the cross-correlation procedure for other fields across the diameter of the artery gives the velocity profile. We modified this procedure: a single field, covering the whole diameter of the vessel was used. This reduces noise to a practicable level and increases the speed of computation; it is also consistent with the one-dimensional formulation of wave intensity. However, the resulting increase in the spread of velocities within each field introduces complexity. Velocity gradients are higher near the wall than near the centre of the vessel, and the pattern of scatterers will therefore be more disrupted near the wall as the region of blood moves down the vessel. As a result, the cross-correlation algorithm is biased towards tracking the scatterers near the centre. Because absolute velocities are higher at the centre, even though velocity gradients are lower, this leads to an overestimate of the mean flow velocity. In our method, this effect was ameliorated by using the centroid rather than the peak of the distribution of correlation coefficients

An additional issue is that the ultrasound image approximates a 2-D longitudinal slice through the centre of the vessel. (In reality, this slice has a finite thickness but that is ignored here.) Considering the simple case of fully developed flow, the velocity profile would be a parabola in the 2-D slice but a paraboloid in the 3-D vessel. The mean velocity for the parabola is three quarters of the maximum velocity, whereas the mean of for the paraboloid is half of the maximum velocity. In the real vessel, where Womersley-type flow occurs, the velocity profile is more complex and it varies over the cardiac cycle. For calculations of wave speed, where absolute rather than relative magnitudes matter, we used the inverse Womersley solution; elsewhere, to minimise computational cost, we assumed Poiseuille flow.

The assessment of arterial diameter over the cardiac cycle also uses the cross-correlation technique but its application in this case is less complex because the wall moves more like a rigid object, with a single velocity at each time step. The method is routinely employed. The movement of the wall can be less than a single pixel; again, this problem arises from the need to use relatively low ultrasound frequencies, but sub-pixel resolution is easily obtained by fitting a curve to the Gaussian cross-correlation function.

The calculation of wave speed from the gradient of the P-U loop or the ln(D)-U loop assumes that all waves are forward-travelling. This condition is most likely to be satisfied during the period when pressure and velocity rise at the start of ventricular ejection. Rabbits and people have broadly similar wave speeds, but the condition is more likely to be violated in the rabbit due to its higher heart rate, meaning there is less time for reflected wave energy to dissipate, and its shorter arteries, which reduce the time before reflections arrive at the measuring site. For these reasons, the present study used the first 10 ms of ejection in rabbits and the first 50 ms in the human subject.

Plots of aortic pressure, obtained with an intra-arterial catheter, and aortic diameter, obtained using the B-mode cross-correlation method (Figure 1), both showed in all animals the expected shape of a rapid systolic rise, followed by a slower decay that was clearly biphasic – that is, there was a secondary, diastolic peak or at least an inflection in the gradient of the decay, before the next systole, and not a continuous exponential decay as might be expected in the absence of wave reflections. The relation between pressure and diameter showed the expected non-linearity and hysteresis, caused by strain-stiffening and viscous effects.

Plots showing blood flow velocities obtained both by the Doppler ultrasound probe mounted at the tip of the intra-arterial catheter and by non-invasive UIV (Figure 2) gave the expected triphasic aortic waveform in all animals and with both techniques: there was a period of forward-going velocity in early systole, followed by a period of nearly zero flow, and finally a period of low, forward going flow in diastole. Furthermore, the shape of the curves obtained by the two methods was similar. A visible discrepancy between the methods was smoothing of the peaks and troughs in the Doppler measurements. The intra-arterial flow measurements were made with a device designed for human use; its sampling rate is probably insufficient to capture fine features of the much shorter cardiac cycle in the rabbit. It also has inbuilt, low-pass filtering. A second discrepancy was a difference in the absolute magnitude in four rabbits: the invasive measurements gave lower peak velocities than were obtained with the non-invasive method. We attribute this to the well-known difficulty in aligning the Doppler probe with the predominant flow direction [11]. With this exception, the scatter plot of the two data sets showed a strong relation, the data lying close to the line of identity.

Wave speeds derived from the ln(D)-U loop were higher than those derived from the P-U loop (Figure 3) but the difference was relatively small: averages were, respectively, 5.36 and 4.48 m/s. As noted above, reflections may have been present in early systole in the rabbit. The presence of reflections introduces errors of opposite direction in the two methods: wave speed is overestimated when using the ln(D)-U loop but under-estimated when using the P-U loop. Comparing data obtained using the two methods with data derived from the ln(D)P loop – thought to be less affected by reflections [30] – supports this view since the latter data lies between the other two data sets (Figure 3). We note that the discrepancies would likely be smaller in people, and also that wave speed is not required when assessing net wave intensity or wave energy (as in the esmolol study), but only for separating waves into their forward and backward components and for investigating vessel wall stiffness.

The P-U and D-U methods gave comparable patterns of wave intensity (Figure 4). In all rabbits, both indicated a large net forward-travelling ("W1") wave in early systole that was followed by a small net backward travelling ("R") wave and then a more complex and dispersed net forward going ("W2") wave. These are thought dominantly to correspond to the initial compression wave caused by ventricular contraction, its reflection, and the forward going expansion wave that derives from an interaction between the inertia of the blood and the reduced and then reversed contraction of the ventricle. The more disperse and complex nature of the W2 wave is thought to be explained at least in part by the longer period between it and the R-wave of the ECG, used to time align individual heart beats when creating an ensemble average. This, coupled with beat-to-beat variation in the duration of the cardiac cycle would have the effect of smearing out a wave even it were concentrated at the same relative location in each individual beat. Additionally, the W2 wave may consist of more components than the other waves. Figure 4 demonstrates a strong correspondence between wave intensities measured by the two methods, and also between wave energies (i.e. the integrals of the intensity curves for each wave). The relation in both cases is curvilinear, as expected from the strain-stiffening behaviour of the wall.

Ventricular dysfunction was transiently induced with esmolol, a short-acting, cardio-selective, $\beta_1$-adrenergic receptor blocker that decreases the force and rate of ventricular contraction. In addition to its clinical use, the drug has been employed in studies investigating the potential performance of left ventricular assist devices in heart failure [e.g. 32, 33]. Both the P-U and D-U methods were unequivocally able to detect decreased intensity, decreased energy and a delayed arrival of the W1 wave at all the esmolol doses employed (Figure 5). Both could also detect similar effects on the energy and arrival time of the W2 wave. The D-U method was less sensitive than the P-U method at detecting the esmolol-induced decrease in peak intensity of the more spread out W2 wave. (Neither method achieved a significant result at the lowest dose.) Effects of viscous properties of the wall on diameter are more important during the W2 than the W1 period; one possibility is that they differ substantially between rabbits, leading to greater variability in the shape of the diameter waveform during late systole and early diastole. In that case, the area under the curve, which is related to the release of stored energy to the blood, would have more consistency than the peak height.

Finally, we speculate on the potential clinical utility of the technique. A proof-of-concept experiment showed that the non-invasive method can be applied to the human carotid artery. The pattern of wave intensities it detected and the calculated wave speed (Figure 6) were consistent with data from other non-invasive techniques [7], Potential uses include screening for heart failure, tracking its progression, and improving the accuracy of prognoses. Although the algorithms required to achieve accurate results have proven complex, they are not difficult to implement now that they have been developed. A recent modelling study [34] suggests that incorporating artificial intelligence into the analysis improves categorisation to the extent that the method might be able to categorise individual patients.

**Acknowledgements**

Supported by MRC Confidence-in-Concept and Proof-of Concept funds, a BHF Research Excellence Award studentship to KR, and British Heart Foundation project grant PG/18/48/33832.

**Disclosures**

PDW is named inventor on patents filed by Imperial College Innovations Ltd that describe some of the underlying technology.

## Supplementary material

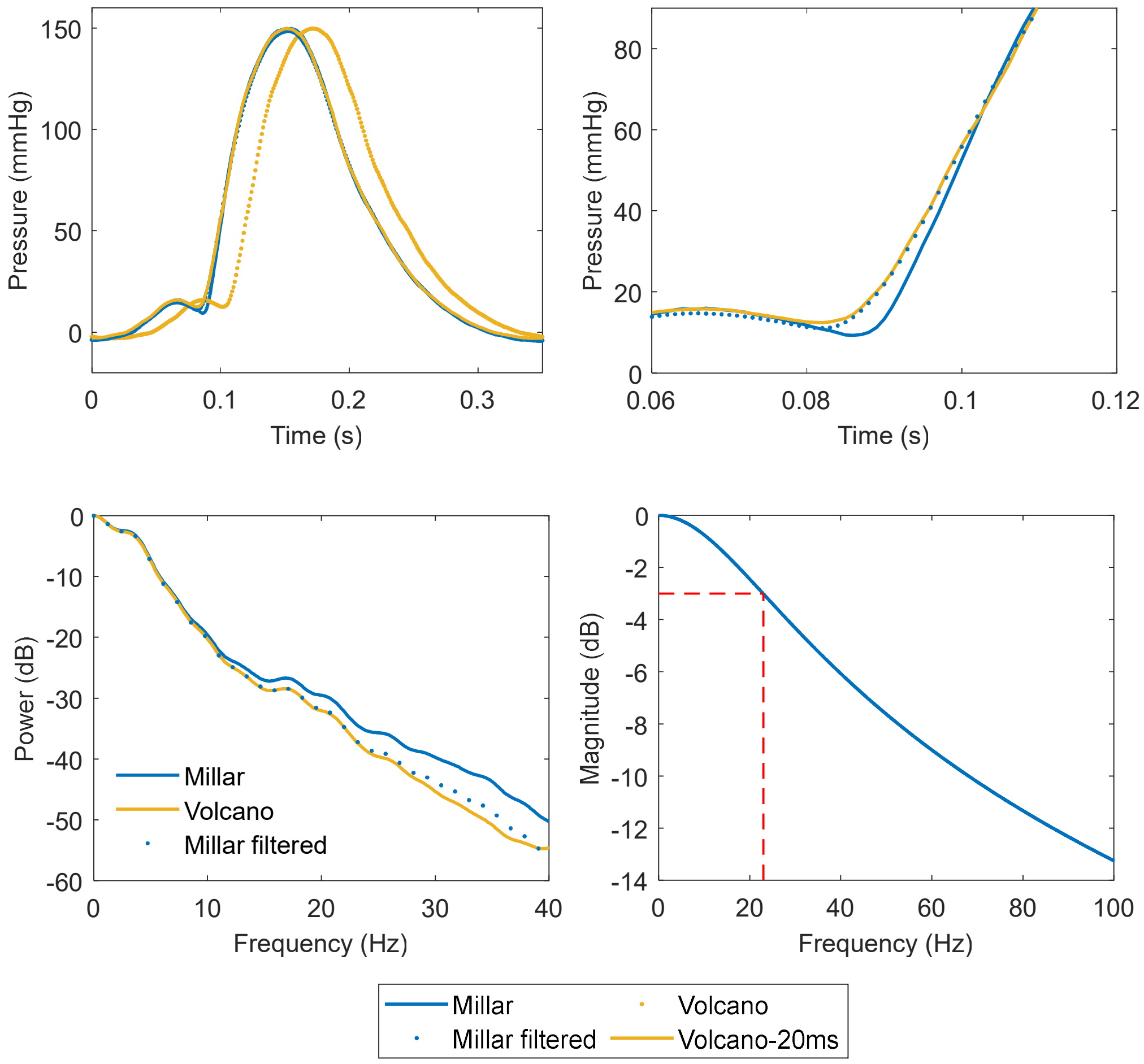

Supplementary Figure 1. (Top left) Pressure data from the Volcano Combowire system (brown points) were compared with data from a high-fidelity Millar catheter (blue line) when both were inserted into a fluid-filled elastic tube connected to a downstream resistor and an upstream pump (Harvard Instruments) that provided a physiologically realistic pulsatile profile at 180 beats/min, typical of anaesthetised rabbits. When the Combowire data are shifted leftwards by 20 ms (brown line), most of the trace coincides with the Millar catheter data but there is a discrepancy at the foot (enlarged in top right panel, brown line vs blue line): the Combowire data are earlier and blunted. Analysis of the frequency spectra (bottom left) shows that the Combowire data are low pass filtered. A first-order, infinite impulse response analog filter with a half power cut-off at 23 Hz (bottom right) gave excellent agreement between Millar and Combowire data when it was applied to the former and the peaks were realigned (brown line vs blue dots, first three panels).

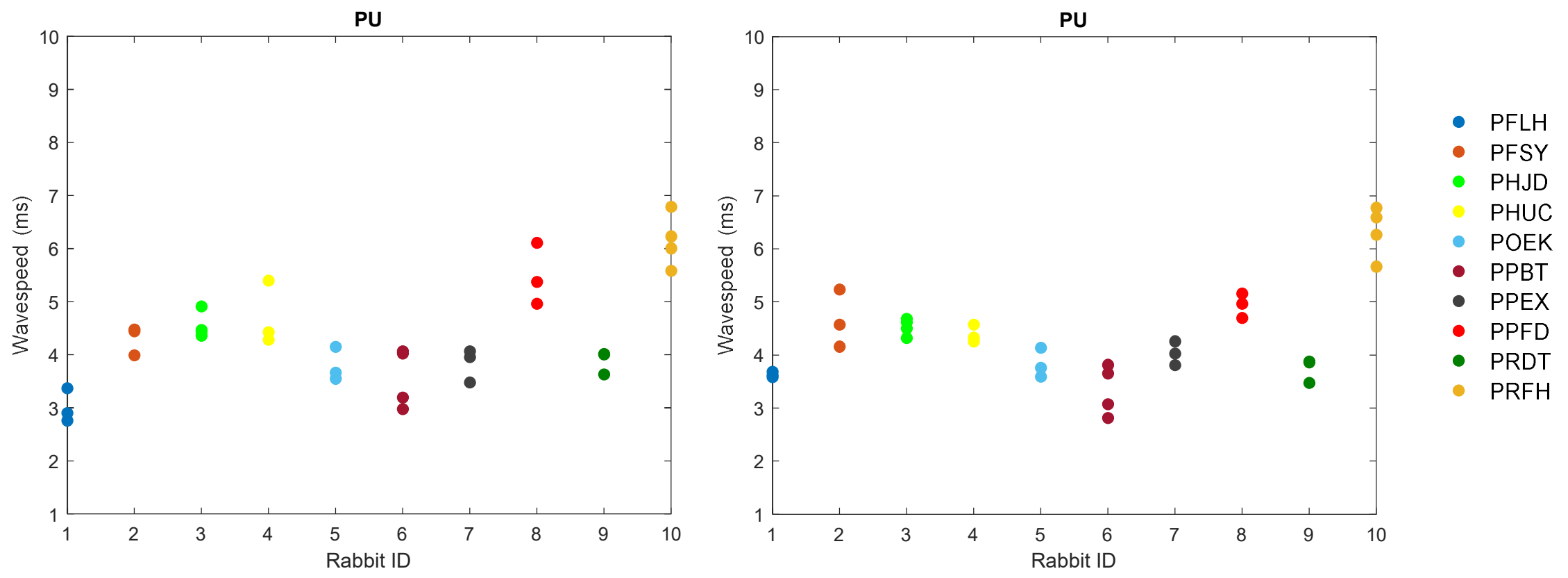

Supplementary Figure 2. Wave speed determined using the PU loop method, showing negligible overall difference between values where U was not (left) or was (right) filtered as described in the text. There is closer grouping of the replicates for most rabbits after filtering.

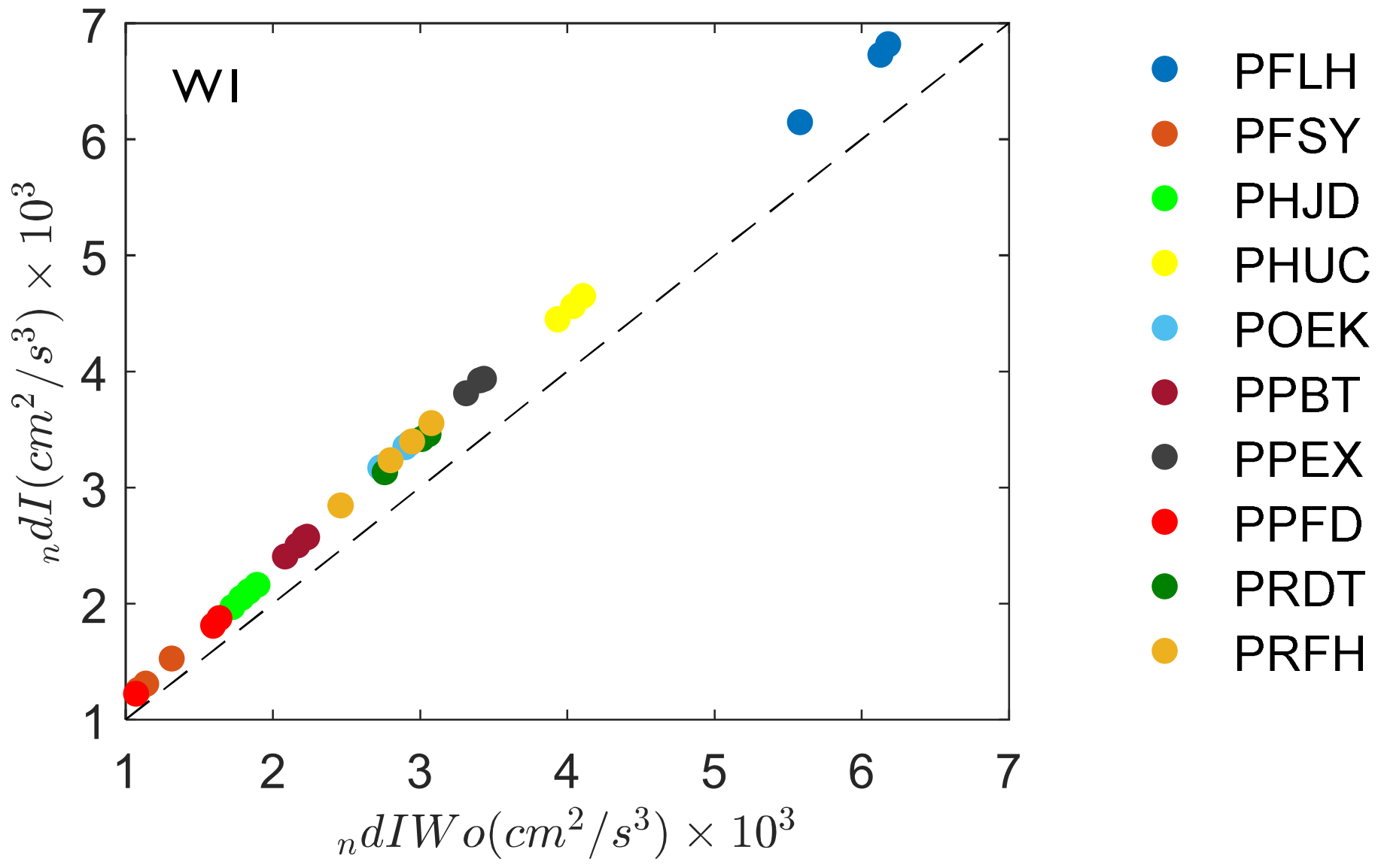

Supplementary Figure 3. Intensity of the W1 wave determined using the 2-D approximation for flow velocity (y-axis) or the full 3-D Womersley solution. Although there is a small quantitative difference in absolute values, there is a strong linear relation between the two estimates.